\documentclass[a4paper,Haag duality]{mathscan}
\newtheorem{thm}{Theorem}[section] 
\newtheorem{pro}[thm]{Proposition}  
\newtheorem{cor}[thm]{Corollary}    
        
\theoremstyle{definition}           
\newtheorem{rem}[thm]{Remark}       
\newtheorem{defn}[thm]{Definition}  

\newcommand{\NI}{\noindent}

\newcommand{\bea}{\begin{eqnarray}}
\newcommand{\eea}{\end{eqnarray}}
\newcommand{\dsp}{\displaystyle}
\def \b #1 {\bf #1}
\newcommand{\IR}{\mathbb{R}}

\newcommand{\IC}{\mathbb{C}}

\newcommand{\IT}{\mathbb{T}}
\newcommand{\IZ}{\mathbb{Z}}
\newcommand{\IM}{\mathbb{M}}

\newcommand{\cal}{\mathcal}
\newcommand{\clk}{{\cal K}}

\newcommand{\cla}{{\cal A}}

\newcommand{\cli}{{\cal I}}

\newcommand{\clh}{{\cal H}}
\newcommand{\clp}{{\cal P}}
\newcommand{\clo}{{\cal O}}
\newcommand{\clb}{{\cal B}}

\newcommand{\clj}{{\cal J}}

\newcommand{\clm}{{\cal M}}

\newcommand{\raro}{\rightarrow}

\newcommand{\vsp}{\vskip 1em}

\newcommand{\ul}{\underline}

\def \qed {\hfill \vrule height6pt width 6pt depth 0pt}
\newcommand{\be}{\begin{equation}}
\newcommand{\ee}{\end{equation}}
\newcommand{\ben}{\begin{eqnarray*}}
\newcommand{\een}{\end{eqnarray*}}
\begin{document}

\title{Spontaneous $SU_2(\IC)$ symmetry breaking in the ground states of quantum spin chain }

\author{ Anilesh Mohari }
\thanks{...}

\address{ The Institute of Mathematical Sciences, CIT Campus, Taramani, Chennai-600113 }

\email{anilesh@imsc.res.in}



\thanks{ I express my gratitude and thanks to anonymous referees of my earlier papers on quantum spin chain for their conspicuous remarks which made it now possible to organize the present paper with finer details. }

\begin{abstract}
In this paper, we have proved that there exists no translation invariant pure state of $\IM=\otimes_{k \in \IZ}\!M^{(k)}_d(\IC)$ that is real, lattice symmetric with a certain twist and $SU_2(\IC)$ invariant for any even integer $d \ge 2$. In particular, this result also says that the Heisenberg iso-spin anti-ferromagnetic model with ${1 \over 2}$-odd integer spin degrees of freedom does not admit a unique ground state.        
\end{abstract}

\maketitle 

\section{ Introduction }

\vsp 
In this paper, we investigate various order properties of ground states for translation invariant Hamiltonian models [BR-II,Sim] in the two-side infinite quantum spin chain $\IM =\otimes_{k \in \IZ}\!M_d^{(k)}(\!C)$ of the following formal form
\be 
H= \sum_{ n \in \IZ} \theta^n(h_0),
\ee
with $h^*_0=h_0 \in \IM_{loc}$, where $\IM_{loc}$ is the union of local sub-algebras of $\IM$ and $\theta$ is the right translation on $\IM$. In particular, our results are aimed to investigate the set of ground states for the Heisenberg anti-ferromagnet iso-spin model $H_{XXX}$ [Be] with nearest neighbour interactions  
\be 
h^{XXX}_0 = J (\sigma_x^0 \otimes \sigma_x^1 +\sigma_y^0 \otimes \sigma_y^1 + \sigma_z^0 \otimes \sigma_z^1),
\ee 
where $\sigma_x^k,\sigma_y^k$ and $\sigma_z^k$ are Pauli spin matrices located at lattice site $k \in \IZ$ and $J > 0$ is a constant. It is well known that any finite volume truncation of $H_{XXX}$ with periodic boundary condition admits a unique ground state [Be,AL]. However, no clear picture has emerged so far in the literature about the set of ground states for the two sided infinite volume anti-ferromagnet Heisenberg $H_{XXX}$ model. However, many interesting results on ground states, those are known for other specific Heisenberg type of models [LSM], such as Ghosh-Majumder (GM) model [GM] and Affleck-Kenedy-Lieb-Tasaki (AKLT) model [AKLT], gave interesting conjectures on the general behaviour of ground states and its physical implication for anti-ferromagnetic Hamiltonian $H_{XXX}$ model. 

\vsp 
One standing conjecture by Haldane [AL] says that $H_{XXX}$ has a unique ground state and the ground state admits a mass gap with its two-point spatial correlation function decaying exponentially for integer spin $s$ ( odd integer $d$, where $d=2s+1$ ). Whereas for the even values of $d$, the conjecture says that $H_{XXX}$ has a unique ground state with no mass gap and its two-point spatial correlation function does not decay exponentially (i.e. $s$ is a ${1 \over 2}$ odd integer spin, where $d=2s+1$). A well known result, due to Affleck and Lieb [AL] says: if $H_{XXX}$ admits a unique ground state for even values of $d$ then the ground state has no mass gap and its two-point spatial correlation function does not decay exponentially. In contrast, if the integer spin $H_{XXX}$ model admits a unique ground state with a mass gap, a recent result [NaS] says that its two point spatial correlation function decays exponentially. Thus the uniqueness of the hypothesis on the ground state for $H_{XXX}$ model is a critical issue to settle a part of the conjecture. We refer interested readers to [AL,Ma3] for finer details on this conjecture and a survey paper [Na] for an overview on this topic. On the experimental side, $H_{XXX}$ finds a special place in the low temperature physics of magnetic materials [Ef,DR] those admit quasi one-dimensional lattice structures.

\vsp 
In this paper, we will use a $C^*$-algebraic method that is independent of Bethe-ansatz [Be] or algebraic Bethe-ansatz [Fa] solution, used in the literature extensively to capture properties of ground states of $H_{XXX}$ model. Nor we will be using the rigorous methods invented in [LSM] and [GM] to study the infinite volume ground states of $H_{XXX}$ as limit points of the finite volume ground states of $H_{XXX}$ with periodic boundary conditions.  

\vsp 
In the following text, we will now formulate the problem in the general framework of $C^*$-dynamical system [BR-II] valid for two-sided one-dimensional quantum spin chain models. The uniformly hyper-finite $C^*$-algebra $\IM=\otimes_{k \in \IZ}\!M_d^{(k)}(\IC)$ of infinite tensor product of $d \times d$-square matrices $\!M_d^{(k)}(\IC) \equiv \!M_d(\IC)$, levelled by $k$ in the lattice $\IZ$ of integers, is the norm closure of the algebraic inductive limit of the net of finite dimensional $C^*$ algebras $\IM_{\Lambda}= \otimes_{k \in \Lambda }\!M_d^{(k)}(\IC)$, where $\Lambda \subset \IZ$ are finite subsets and an element $Q$ in $\IM_{\Lambda_1}$ is identified with the element $Q 
\otimes I_{\Lambda_2 \bigcap \Lambda_1^c}$ in $\IM_{\Lambda_2}$, i.e. by the inclusion map if $\Lambda_1 \subseteq \Lambda_2$, where $\Lambda^c$ is the complementary set of $\Lambda$ in $\IZ$.
We will use the symbol $\IM_{loc}$ to denote the union of all local algebras $\{ \IM_{\Lambda}: \Lambda \subset \IZ,\;|\Lambda| < \infty \}$. Thus $\IM$ is a quasi-local $C^*$-algebra with 
local algebras $\{\IM_{\Lambda}:|\Lambda| < \infty \}$ and $\IM_{\Lambda}'=\IM_{\Lambda^c}$, where $\IM'_{\Lambda}$ is the commutant of $\IM_{\Lambda}$ in $\IM$. We refer readers to Chapter 6 of [BR-II] for more details on quasi-local $C^*$-algebras. 

\vsp 
The lattice $\IZ$ is a group under addition and for each $n \in \IZ$, we have an automorphism $\theta^n$, extending the translation action, which takes $Q^{(k)}$ to $Q^{(k+n)}$ for any $Q \in \!M_d(\IC)$ and $k \in \IZ$, by the linearity and multiplicative properties on $\IM$. A unital positive linear functional $\omega$ of $\IM$ is called {\it state}. It is called {\it translation-invariant} if $\omega = \omega \theta$. A linear automorphism or anti-automorphism $\beta$ [Ka] on $\IM$ is called {\it symmetry } for $\omega$ if $\omega \beta = \omega$. Our primary objective is to study translation-invariant states and their symmetries that find relevance in Hamiltonian dynamics of quantum spin chain models $H$ [BR-II,Ru,Sim].      

\vsp 
We consider [BR-II,Chapter 6],[Ru] quantum spin chain Hamiltonian in one dimensional lattice $\IM$ of the following form
\be 
H= \sum_{ n \in \IZ} \theta^n(h_0)
\ee
for $h^*_0=h_0 \in \IM_{loc}$, where the formal sum in (3) gives a group of auto-morphisms $\alpha=(\alpha_t:t \in \IR)$ by 
the thermodynamic limit: $\mbox{lim}_{\Lambda_{\eta} \uparrow \IZ}||\alpha^{\Lambda_{\eta}}_t(A)-\alpha_t(A)||=0$ for all $A \in \IM$ and $t \in \IR$ for 
a net of finite subsets $\Lambda_{\eta}$ of $\IZ$ with uniformly bounded surface energy, where automorphisms $\alpha^{\Lambda}_t(x)=e^{itH_{\Lambda}}xe^{-itH_{\Lambda}}$ is determined by the finite subset $\Lambda$ of $\IZ^k$ and $H_{\Lambda}=\sum_{n \in \Lambda} \theta^n(h_0)$. Furthermore, the limiting automorphism $(\alpha_t)$ does not depend on the net that we choose in the thermodynamic limit $\Lambda_{\eta} \uparrow \IZ$, provided the surface energies of $H_{\Lambda_\eta}$ are kept uniformly bounded. The uniquely determined group of automorphisms $(\alpha_t)$ on $\IM$ is called {\it Heisenberg flows } of $H$. In particular,  we have $\alpha_t \circ \theta^n = \theta^n \circ \alpha_t$ for all $t \in \IR$ and $n \in \IZ$. Any linear automorphism or anti-automorphism $\beta$ on $\IM_{loc}$, keeping the formal sum (3) in $H$ invariant, will also commute with $(\alpha_t)$.  

\vsp 
A state $\omega$ is called {\it stationary} for $H$ if $\omega \alpha_t= \omega$ on $\IM$ for all $t \in \IR$. The set of stationary states of $H$ is a non-empty compact convex set and has been extensively studied in the last few decades within the framework of ergodic theory for $C^*$-dynamical systems [BR-I,Chapter 4]. However, a stationary state of $H$ need not be always translation-invariant. A stationary state $\omega$ of $\IM$ for $H$ is called $\beta$-KMS state at an inverse positive temperature $\beta > 0$ if there exists a function $z \raro f_{A,B}(z)$, analytic on the open strip $0 < Im(z) < \beta$, bounded continuous on the closed strip $0 \le Im(z) \le \beta$ with boundary condition 
$$f_{A,B}(t)=\omega_{\beta}(\alpha_t(A)B),\;\;f_{A,B}(t+i\beta)=\omega_{\beta}(\alpha_t(B)A)$$
for all $A,B \in \IM$. Using weak$^*$ compactness of convex set of states on $\clm$, finite volume Gibbs state $\omega_{\beta,\Lambda}$ is used to prove existence of a KMS state $\omega_{\beta}$ for $(\alpha_t)$ at inverse positive temperature $\beta > 0$. 
The set of KMS states of $H$ at a given inverse positive temperature $\beta$ is singleton set i.e. there is a unique $\beta$ KMS-state at a given inverse positive temperature $\beta={ 1 \over kT }$ for $H$ which has a finite range interaction [Ara1],[Ara2], [Ki] and thus inherits translation and other symmetry of the Hamiltonian. The unique KMS states of $H$ at a given inverse temperature is ergodic for translation dynamics. This gives a strong motivation to study translation-invariant states in a more general framework of $C^*$-dynamical systems [BR-I].    

\vsp 
A state $\omega$ of $\IM$ is called {\it ground state} for $H$, if the following two conditions are satisfied:

\NI (a) $\omega(\alpha_t(A))=\omega(A)$ for all $t \in \IR$; 

\NI (b) If we write on the GNS space $(\clh_{\omega},\pi_{\omega},\zeta_{\omega})$ of $(\IM,\omega)$, $$\alpha_t(\pi_{\omega}(A))=e^{itH_{\omega}}\pi_{\omega}(A)e^{ -itH_{\omega}}$$ 
for all $A \in \IM$ with $H_{\omega}\zeta_{\omega}=0$, then $H_{\omega} \ge 0$.   

\vsp 
Furthermore, we say a ground state $\omega$ is {\it non-degenerate}, if null space of $H_{\omega}$ is spanned by $\zeta_{\omega}$ only. We say $\omega$ has a {\it mass gap}, if the spectrum $\sigma(H_{\omega})$ of $H_{\omega}$ is a subset of $\{ 0 \} \bigcap [\delta, \infty)$ for some $\delta >0$. For a wide class of spin chain models [NaS], which includes  Hamiltonian $H$ with finite range interaction, $h_0$ being in $\IM_{loc}$, the existence of a non vanishing spectral gap of a ground state $\omega$ of $H$ implies exponential decaying two-point spatial correlation functions. We present now a precise definition for exponential decay of two-point spatial correlation functions of a state $\omega$ of $\IM$. We use symbol $\Lambda^c_m$ for complementary set of the finite volume box $\Lambda_m = \{ n: -m \le n \le m \}$ for $m \ge 1$. 

\vsp
\begin{defn} 
Let $\omega$ be a translation-invariant state of $\IM$. We say that the two-point spatial correlation functions of $\omega$ {\it decay exponentially}, if there exists a $\delta > 0$ satisfying the following condition: for any two local elements $Q_1,Q_2 \in \IM$ and $\epsilon > 0$, there exists an integer $m \ge 1$ such that    
\be
e^{\delta |n|} |\omega( Q_1 \theta^n(Q_2) ) - \omega(Q_1) \omega(Q_2)| \le \epsilon
\ee
for all $n \in \Lambda^c_m$.  
\end{defn}

\vsp 
By taking low temperature limit of $\omega_{\beta}$ as $\beta \raro \infty$, one also proves existence of a ground state for $H$ [Ru,BR-II]. On the contrary to KMS states, the set of ground states is a convex face in the set of the convex set of $(\alpha_t)$ invariant states of $\IM$ and its extreme points are {\it pure} states of $\IM$ i.e. A state is called {\it pure} if it can not be expressed as convex combination of two different states of $\IM$. Thus low temperature limit points of unique $\beta-$KMS states give ground states for the Hamiltonian $H$ inheriting translation and other symmetry of the Hamiltonian. In general the set of ground states need not be a singleton set and there could be other states those are not translation invariant but still a ground state for a translation invariant Hamiltonian. Ising model admits non translation invariant ground states known as N\'{e}el state [BR vol-II]. However ground states that appear as low temperature limit of $\beta-$KMS states of a translation invariant Hamiltonian, inherit translation and other symmetry of the Hamiltonian. In particular if ground state of a translation invariant Hamiltonian model (3) is unique, then the ground state is a translation invariant pure state. 

\vsp
Let $Q \raro \tilde{Q}$ be the automorphism on $\IM$ that maps an element 
$$Q=Q_{-l}^{(-l)} \otimes Q_{-l+1}^{(-l+1)} \otimes ... \otimes Q_{-1}^{(-1)} \otimes Q_0^{(0)} \otimes Q_1^{(1)} ... \otimes Q_n^{(n)}$$ by reflecting around the point ${1 \over 2}$ of the lattice $\IZ$ to 
$$\tilde{Q}= Q_n^{(-n+1)}... \otimes Q_1^{(0)} \otimes Q_0^{(1)} \otimes Q_{-1}^{(2)} \otimes ... Q_{-l+1}^{(l)} \otimes Q_{-l}^{(l+1)}$$
for all $n,l \ge 1$ and $Q_{-l},..Q_{-1},Q_0,Q_1,..,Q_n \in 
M_d(\IC)$. 

\vsp 
For a state $\omega$ of $\IM$, we set a state $\tilde{\omega}$ of $\IM$ by 
\be 
\tilde{\omega}(Q)= \omega(\tilde{Q})
\ee
for all $Q \in \IM$. Thus $\omega \raro \tilde{\omega}$ is an affine one to one onto 
map on the convex set of states of $\IM$. The state $\tilde{\omega}$ is translation-invariant if and only if $\omega$ is translation-invariant state. We say a state $\omega$ is {\it lattice reflection-symmetric} or in short {\it lattice symmetric } if $\omega=\tilde{\omega}$.   

\vsp 
The group of unitary matrices $u \in U_d(\IC)$ acts naturally on $\IM$ as a group of automorphisms of $\IM$ defined by 
\be 
\beta_{u}(Q)=(..\otimes u \otimes u \otimes ...)Q(...\otimes u^* \otimes u^* \otimes u^*...)
\ee
We also set automorphism $\tilde{\beta}_u$ on $\IM$ defined by 
\be 
\tilde{\beta}_u(Q)=\beta_u(\tilde{Q})
\ee
for all $Q \in \IM$. So for $u,w \in U_d(\IC)$, we have 
$$\tilde{\beta}_u \tilde{\beta}_w=\beta_{uw}$$
In particular, $\tilde{\alpha}_{w}^2(Q)=Q$ for all $Q \in \IM$ if and only if $w^2=I_d$. We say a state $\omega$ of $\IM$ is {\it lattice symmetric with a twist } $w \in U_d(\IC)$ if 
\be 
w^2=I_d,\;\;\omega(\tilde{\beta}_{w}(Q))=\omega(Q)
\ee

\vsp 
We fix an orthonormal basis $e=(e_i)$ of $\!C^d$ and $Q^t \in \!M_d(\IC)$ be the transpose of $Q \in \!M_d(\IC)$ with respect to an orthonormal basis $(e_i)$ for $\IC^d$ (not complex conjugate). Let $Q \raro Q^t$ be the linear anti-automorphism map on $\IM$ that takes an element 
$$Q= Q^{(l)}_0 \otimes Q^{(l+1)}_1 \otimes ....\otimes Q^{(l+m)}_m$$
to its transpose with respect to the basis $e=(e_i)$ defined by
$$Q^t={Q^t_0}^{(l)} \otimes {Q^t_1}^{(l+1)} \otimes ..\otimes {Q^t_m}^{(l+m)},$$
where $Q_0,Q_1,...,Q_m$ are arbitrary elements in $\!M_d(\IC)$. We also note that $Q^t$ depends on the basis $e$ that we choose and we have avoided use of a suffix $e$. He assumed that it won't confuse an attentive reader since we have fixed an orthonormal basis $(e_i)$ for our consideration through out this paper. For more general $Q \in \IM_{loc}$, we define $Q^t$ by extending linearly and take the unique bounded linear extension for any $Q \in \IM$. For a state $\omega$ of $\IM$, we define a state $\bar{\omega}$ on $\IM$ by the following prescription
\be
\bar{\omega}(Q) = \omega(Q^t)
\ee
Thus the state $\bar{\omega}$ is translation-invariant if and only if $\omega$ is translation-invariant. We say $\omega$ is {\it real }, if $\bar{\omega}=\omega$. The formal Hamiltonian $H$ is called {\it reflection symmetric with twist $w$ } if $\beta_{w}(\tilde{H})=H$ and {\it real} if $H^t=H$. 

\vsp 
We also set a conjugate linear map $Q \raro \overline{Q}$ on $\IM$ with respect to the basis $(e_i)$ for $\IC^d$ defined by extending the identity action on elements  
$$..I_d \otimes |e_{i_0}\rangle \langle e_{j_0}|^{(k)} \otimes |e_{i_1}\rangle \langle e_{j_1}|^{(k+1)} \otimes 
|e_{i_n}\rangle \langle e_{j_n}|^{(k+n)} \otimes I_d ..,\;1 \le i_k,j_k \le d,\;\;k \in \IZ,\;n \ge 0$$ 
anti-linearly. Thus by our definition we have 
$$Q^*=\overline{Q^t}$$ 
and 
$$(\overline{Q})^*=\overline{Q^*}$$

\vsp 
We set the following anti-linear reflection map $\clj_{w}:\IM \raro \IM$ with twist $w \in U_d(\IC)$, defined by 
\be 
\clj_{w}(Q) = \overline{\beta_{w}(\tilde{Q})}
\ee 
for all $Q \in \IM$. 

\vsp 
Following a well known notion [FILS], a state $\omega$ on $\IM$ is called {\it reflection positive with a twist $r_0 \in U_d(\IC),\;r_0^2=I_d$}, if
\be 
\omega(\clj_{r_0}(Q) Q) \ge 0
\ee 
for all $Q \in \IM_R$. Thus the notion of reflection positivity also depends explicitly on the underlining fixed orthonormal basis $e=(e_i)$ of $\IC^d$.

\vsp 
Let $G$ be a compact group and $g \raro u(g)$ be a $d-$dimensional unitary representation of $G$. By $\gamma_g$ we denote the product action of $G$ on the infinite tensor product $\IM$ induced by $u(g)$,
\be 
\gamma_g(Q)=(..\otimes u(g) \otimes u(g)\otimes u(g)...)Q(...\otimes u(g)^*\otimes u(g)^*\otimes u(g)^*...)
\ee
for any $Q \in \IM$, i.e. $\gamma_g=\beta_{u(g)}$. We say $\omega$ is $G$-invariant, if 
\be 
\omega(\gamma_g(Q))=\omega(Q)
\ee 
for all $Q \in \IM_{loc}$. If $G=U_d(\IC)$ and $u:U_d(\IC) \raro U_d(\IC)$ is the natural representation $u(g)=g$, then we will identify the notation $\beta_g$ with $\gamma_g$ for 
simplicity. Formal Hamiltonian $H$ given in (3) is called $G$-gauge invariant if $\gamma_g(H)=H$
for all $g \in G$.  

\vsp 
We recall now [DLS,FILS] if $H$ in (3) has the following form 
\be 
-H= B + \clj_{r_0}(B) + \sum_i C_i \clj_{r_0}(C_i)
\ee 
for some $B, C_i \in \IM_R$ then the unique KMS state at inverse positive temperature $\beta$ is refection positive with the twist $r_0$. We refer to [FILS] for details, which we will cite frequently while dealing with examples satisfying (12). Since the weak$^*$-limit of a sequence of reflection positive states with the twist $r_0$ is also a reflection positive state with the twist $r_0$, weak$^*$-limit points of the unique $\beta-$KMS state of $H$ as $\beta \raro \infty$, are also refection positive with the twist $r_0$. Thus any weak$^*$ low temperature limit point ground state of $H$ is reflection positive with a twist $r_0$ if 
$H$ is given by (12). 

\vsp 
In particular, the anti-ferromagnetic $H_{XXX}$ models are real and reflection symmetric admitting the form (12) [FILS] with $r_0=\sigma_y$. 
Another mathematically instructive example of real reflection symmetric Hamiltonian satisfying (12) is the anti-ferro-magnetic $XY$ model $H_{XY}$ defined by   
\be 
h^{XY}_0 = J (\sigma_x^0 \otimes \sigma_x^1 + \sigma_y^0 \otimes \sigma_y^1)
\ee 
for some constant $J > 0$. The model $H_{XY}$ is exactly solvable [LSM,AMa] and its partition function at positive temperatures [LSM] are known explicitly. Furthermore, it is also well known that $H_{XY}$ admits a unique ground state [AMa] and the ground state admits no mass gap [AL]. Furthermore, its two-point spatial correlation function does not decay exponentially [Ma3]. We will get back to this model in the last section of this paper with some additional results for its unique ground state.   

\vsp 
Hamiltonian $H_{XXX}$ admits $SU_2(\IC)$ gauge symmetry with irreducible representation $g \raro u(g)$.  Whereas $H_{XY}$ admits $S^1 \subset SU_2(\IC)$ gauge symmetry, where an element $z \in S^1$ is identified to 
the following element in $SU_2(\IC)$: 
\be
g_z = \left (\begin{array}{llll} z&,&\;\; 0 \\ 0&,&\;\;\bar{z}
\end{array} \right ). 
\ee

\vsp 
A pure mathematical question that arise here: does this additional symmetry of $H$ helps to understand behaviour of its low temperature limiting ground states? Taku Matsui had investigated [Ma3] this question for translation invariant pure states $\omega$ of $\IM=\otimes_{k \in \IZ} \!M^{(k)}_d(\IC)$. In particular, we proved that one of the following statements is false for even integer $d$:

\NI (a) $\pi_{\omega}(\IM_R)''$ is a type-I factor state;  

\NI (b) $\omega$ is $SU_2(\IC)$ gauge invariant with an irreducible representation $g \raro u(g)$. 

In general, for a translation invariant pure state $\omega$, $\pi_{\omega}(\IM_R)''$ need not be a type-I factor [Ma3]. However, it is known that $\pi_{\omega}(\IM_R)''$ is either a type-I 
or a type-III factor [Mo1, Mo3]. 

\vsp 
As an application of our main mathematical results of this paper, we will prove the following theorem in the fourth section. 

\vsp 
\begin{thm} 
Let $\omega$ be a translation invariant, real, reflection positive with twist $r_0 \in U_d(\IC)$ state of $\IM=\otimes_{k \in \IZ}\!M^{(k)}_d(\IC)$. Then at least one of the following two statements is not true for even values of $d$:

\NI (a) $\omega$ is pure;

\NI (b) $\omega$ is $SU_2(\IC)$-invariant, where $g \raro u(g) \in U_d(\IC)$ in (13) is an irreducible representation of $SU_2(\IC)$ satisfying 
\be 
r_0^2=I_d,\;\; r_0u(g)r_0^*=\bar{u(g)}
\ee
for all $g \in SU_2(\IC)$, where the matrix conjugation with respect to an orthonormal basis $e=(e_i)$ of $\IC^d$. 
\end{thm}    

\vsp 
As an application of Theorem 1.2, we will prove in the last section the following corollary.

\vsp 
\begin{cor} 
Let $H$ be a translation invariant Hamiltonian of the form $H=\sum_{ k \in \IZ} \theta_k(h_0)$ with $h_0=h_0^* \in \IM_{loc}$. Let $H$ be also $SU_2(\IC)$ invariant with an irreducible representation $g \raro u(g)$ of $SU_2(\IC)$ and $r_0$ be the element in $U_d(\IC)$ satisfying (17). 
Let $H$ be also real (with respect to the basis $e=(e_i)$ ), lattice reflection symmetric and unique $\beta$-KMS at inverse positive temperature be reflection symmetric with the twist $r_0 \in U_d(\IC)$. If $d$ is an even integer, then the set of ground states for $H$ is not singleton.       
\end{cor} 

\vsp 
However, we have the following important corollaries. 

\vsp 
\begin{cor} 
Let $\omega$ be a translation invariant ground state for $H$ given in Corollary 1.3 that is real, lattice-symmetric and $SU_2(\IC)$ invariant with an irreducible representation $g \raro u(g)$ of $SU_2(\IC)$. If $d$ is an even integer and the ground state $\omega$ is non-degenerate then the following hold:

\NI (a) $\omega$ is not a factor state; 

\NI (b) $H_{\omega}$ has no mass gap.

\end{cor} 

\vsp 
\begin{cor} 
Let $H_{XXX}$ be the Heisenberg ${1 \over 2}$-odd integer anti-ferromagnetic spin model in quantum spin chain $\IM= \otimes_{k \in \IZ} \!M^{(k)}_d(\!C)$, $d$ is an even integer. Then 
the following hold:

\NI (a) Any ground state of $H_{XXX}$ that is a low temperature limit of thermal equilibrium states is not pure. 

\NI (b) The model $H_{XXX}$ does not have a unique ground state. 

\NI (c) Any infinite volume thermodynamic limit of finite volume Bethe states with periodic condition is translation invariant and refection positive with the twist $r_0$ but not pure on $\IM$. Such a ground state has no mass gap if it is non-generate.  
\end{cor}  

\vsp  
Thus our analysis finally gives a surprising result for ${1 \over 2}$ odd integer spin anti-ferromagnetic $H_{XXX}$ contrary to general belief for the last few decades. However it does not rule out possibility of a unique limit point while taking low temperature limit $T \raro 0$ and thus also does not rule out possibility of a strongly correlated two-point spatial correlation function for its low temperature limiting ground state as per assertion of Corollary 1.4. Thus one important question that remains to be answered whether ground state of integer spin $H_{XXX}$ model is unique?   

\vsp 
The paper is organized as follows: In section 2, we will recall basic mathematical set up required from earlier paper [Mo3] and explain basic ideas involved in the proof of Theorem 1.2. In section 4 we give the proof of Theorem 1.2. In the last section, we will illustrate our results with models of physical interest. In particular, we will give proofs of Corollary 1.3, Corollary 1.4 and Corollary 1.5.       
      
\section{Mathematical Preliminaries} 

\vsp 
For the last few decades, a translation-invariant state of $\IM$ had been studied extensively in the mathematical literature, either in the framework of quantum Markov states [Ac], [FNW1], [FNW2], [FNW3] or in the frame work of representation theory of $C^*$ algebras  [Pow], [Cu], [BJ], [BJP] and [BJKW]. Our investigation in [Mo2] and [Mo3] had clubbed these two frameworks into a unified Kolmogorov dilation theory [Mo1], where inductive limit states [Sa] are visualized in the frame work of Kolmogorov consistency theorem for stationary Markov processes. In this section, however, we give the basic ideas that are involved in the proof of Theorem 1.2 after recalling some known results from [BJKW], [Mo2] and [Mo3] for our present purpose.  

\vsp 
A state $\omega$ on a $C^*$-algebra $\IM$ is called factor, if the center of the von-Neumann algebra $\pi_{\omega}(\IM)''$ is trivial, where $(\clh_{\omega},\pi_{\omega},\zeta_{\omega})$ is the Gelfand-Naimark-Segal (GNS) space associated with $\omega$ on $\IM$ [BR-I] and $\pi_{\omega}(\IM)''$ is the double commutant of $\pi_{\omega}(\IM)$. A state $\omega$ on $\IM$ is pure, if $\pi_{\omega}(\IM)''=\clb(\clh_{\omega})$, the algebra of all bounded operators on $\clh_{\omega}$. Here we fix our convention that Hilbert spaces that are considered here are always equipped with inner products $ \langle .,. \rangle $ which are linear in the second variable and conjugate linear in the first variable.
We also recall [Theorem 2.7 in [Pow] or Corollary 2.6.11 in [BR-I], [Ma3] a standard definition of a state to be split in the following. 

\vsp 
Let $\omega$ be a translation-invariant state of $\IM$ and $\omega_{\Lambda}$ be the state $\omega$ restricted to $\IM_{\Lambda}$. We say that $\omega$ is {\it split}, if the following condition is valid for any subset $\Lambda$ of $\IZ$: Given any $\epsilon > 0$ there exists a $m \ge 1$, so that
\be
\mbox{sup}_{||Q|| \le 1}|\omega(Q)-\omega_{\Lambda} \otimes \omega_{\Lambda^c}(Q)| \le \epsilon,
\ee
where the above sup is taken over all local elements $Q \in \IM_{ \Lambda^c_m }$ with the norm less than equal to $1$. The uniform clustering property (17) of the state $\omega$ has its mathematical appeal which guarantees that $\omega$ is {\it quasi equivalent } to the tensor product state $\omega_{\Lambda} \otimes \omega_{\Lambda^c}$ by Theorem 2.7 in [Pow]. A {\it Gibbs state } [BR-II, Chapter 6.2.2] of a Hamiltonian with finite range interaction is split. In particular, if $\omega$ is pure and split then $\omega_R$ is a type-I factor state. However, a pure state need not be a split state [Ma3]. 

\vsp 
We recall in the following, a well known result (Theorem 2.5 in [Pow]). A translation-invariant state $\omega$ of $\IM$ is a factor state if and only if the following holds: for any given $Q_2 \in \IM$ and $\epsilon > 0$, there exists an integer $n \ge 1$ so that  
\be 
\mbox{sup}_{ Q_1 \in \IM_{\Lambda^c_n},||Q_1|| \le 1}|\omega(Q_1 Q_2) - \omega(Q_1)\omega(Q_2)| \le \epsilon 
\ee 
The criteria givin in (18) is used to deduce that a translation-invariant state $\omega$ of $\IM$ is a factor state, if and only if $\omega_{\Lambda}$ ( $\omega_{\Lambda^c}$ ) is a factor state for all subsets of $\Lambda$ of $\IZ$. 

\vsp
We recall that the Cuntz algebra $\clo_d ( d \in \{2,3,.., \} )$ [Cun] is the universal unital $C^*$-algebra generated by the elements $\{s_1,s_2,...,
s_d \}$ subjected to the following relations:
\be 
s_i^*s_j = \delta^i_j I,\;\;\sum_{1 \le i \le d } s_is^*_i=I
\ee

\vsp
Let $\Omega=\{1,2,3,...,d\}$ be a set of $d$ elements. $\cli$ be the set of finite sequences
$I=(i_1,i_2,...,i_m)$ of elements, where $i_k \in \Omega$ and $m \ge 1$ and we use notation 
$|I|$ for the cardinality of $I$. We also include null set denoted by $\emptyset$ in the collection $\cli$ and set $s_{\emptyset }=s^*_{\emptyset}=I$ identity of $\clo_d$ and $s_{I}=s_{i_1}......s_{i_m} \in \clo_d $ and $s^*_{I}=s^*_{i_m}...s^*_{i_1} \in \clo_d$. 

\vsp
The group $U_d(\IC)$ of $d \times d$ unitary matrices acts canonically on $\clo_d$ as follows:
$$\beta_u(s_i)=\sum_{1 \le j \le d}\overline{u^j_i}s_j$$
for $u=((u^i_j) \in U_d(\IC)$. In particular,  the gauge action is defined by
$$\beta_z(s_i)=zs_i,\;\;z \in \IT = S^1= \{z \in \IC: |z|=1 \}.$$
The fixed point sub-algebra of $\clo_d$ under the gauge action i.e., 
$\{x \in \clo_d: \beta_z(x)=x,\;z \in S^1 \}$ is the closure of 
the linear span of all Wick ordered monomials of the form
\be 
s_{i_1}...s_{i_k}s^*_{j_k}...s^*_{j_1}:\;I=(i_1,..,i_k),J=(j_1,j_2,..,j_k)
\ee
and is isomorphic to the uniformly hyper-finite $C^*$ sub-algebra
$$\IM_R =\otimes_{1 \le k < \infty}\!M^{(k)}_d(\IC)$$
of $\IM$, where the isomorphism carries the Wick ordered monomial (20) 
into the following matrix element 
\be 
|e^{i_1}\rangle \langle e_{j_1}|^{(1)} \otimes |e^{i_2}\rangle\langle e_{j_2}|^{(2)} \otimes....\otimes |e^{i_k}\rangle\langle e_{j_k}|^{(k)} \otimes 1 \otimes 1 ....
\ee
We use notation $\mbox{UHF}_d$ for the fixed point $C^*$ sub-algebra of $\clo_d$ under the gauge group action 
$(\beta_z:z \in S^1)$. The restriction of $\beta_u$ to $\mbox{UHF}_d$ is then carried into action
$$Ad(u)\otimes Ad(u) \otimes Ad(u) \otimes ....$$
on $\IM_R$.

\vsp
We also define the canonical endomorphism $\lambda$ on $\clo_d$ by
\be 
\lambda(x)=\sum_{1 \le i \le d}s_ixs^*_i
\ee
and the isomorphism carries $\lambda$ restricted to $\mbox{UHF}_d$ into the one-sided shift
$$y_1 \otimes y_2 \otimes ... \raro 1 \otimes y_1 \otimes y_2 ....$$
on $\IM_R$. We note for all $u \in U_d(\IC)$ that $\lambda \beta_u = \beta_u \lambda$ 
on $\clo_d$ and so in particular, also on $\mbox{UHF}_d$. 

\vsp 
A family $(v_k:1 \le k \le d)$ of contractive operators on a Hilbert space $\clk$ is called {\it a Popescu element} [Po], if 
\be 
\sum_k v_kv_k^*=I_{\clk}
\ee
For a Popescu element $(v_k:1 \le k \le d)$ on a Hilbert space $\clk$, we define a unital completely positive map $\tau$ on $\clb(\clk)$ by 
\be 
\tau(x)=\sum_k v_k x v_k^*,\;x \in \clb(\clk)
\ee
and $\tau$-invariant elements $\clb_{\tau}(\clk)$ in $\clb(\clk)$ by 
\be 
\clb(\clk)_{\tau}=\{ x \in \clb(\clk): \tau(x)=x \}
\ee 
We also note that the group action $(\beta_u)$ of $U_d(\IC)$ on the collection of Popescu elements $(v_i)$ defined by 
$$\beta_u(v_i)= \sum_{1 \le j \le d}\overline{u^j_i}v_j,\;\;1 \le j \le d$$
keeps $\clb(\clk)_{\tau}$ unperturbed. 

\vsp 
We recall Proposition 2.4 in [Mo3] with little more details in the following proposition. The proof given for Proposition 2.4 (a) is valid for any $\lambda$-invariant state of $\clo_d$.  

\vsp 
\begin{pro} 
Let $(\clh_{\psi},\pi_{\psi},\zeta_{\psi})$ be the GNS representation of a $\lambda$ invariant state $\psi$ on $\clo_d$ and $P$ be the support projection of the normal state $\psi_{\zeta_{\psi}}(X)=\langle\zeta_{\psi},X\zeta_{\psi}\rangle$ in the 
von-Neumann algebra $\pi_{\psi}(\clo_d)''$. Then the following holds:

\vsp
\NI (a) $P$ is a sub-harmonic projection for the endomorphism $\Lambda(X)=\sum_k S_kXS^*_k$ on $\pi_{\psi}(\clo_d)''$
i.e. $\Lambda(P) \ge P$ satisfying the following:

\NI (i) $PS^*_kP=S^*_kP,\;\;1 \le k \le d$;

\NI (ii) The set $\{ S_If: Pf=f,\;f \in \clh_{\psi}, |I| < \infty \}$ is total in $\clh_{\psi}$; 

\NI (iii) $\Lambda_n(P) \uparrow I$ as $n \uparrow \infty$;

\NI (iv) $\sum_{1 \le k \le d} v_kv_k^*=I_{\clk};$ 

where $S_k=\pi_{\psi}(s_k)$ and $v_k=PS_kP$ for $1 \le k \le d$ are contractive operator on Hilbert subspace $\clk$, the range of the projection $P$;

\vsp
\NI (b) For any $I=(i_1,i_2,...,i_k),J=(j_1,j_2,...,j_l)$ with $|I|,|J| < \infty$ we have $\psi(s_Is^*_J) =
\langle \zeta_{\psi},v_Iv^*_J\zeta_{\psi}\rangle$ and the vectors $\{ S_If: f \in \clk,\;|I| < \infty \}$ are total in $\clh_{\psi}$;

\vsp
\NI (c) The von-Neumann algebra $\clm=P\pi_{\psi}(\clo_d)''P$, acting on the Hilbert space
$\clk$ i.e. range of $P$, is generated by $\{v_k,v^*_k:1 \le k \le d \}''$ and the normal state
$\phi(x)=\langle\zeta_{\psi},x \zeta_{\psi}\rangle$ is faithful on the von-Neumann algebra $\clm$.

\vsp 
\NI (d) The linear map $X' \in \pi_{\psi}(\clo_d)'$ is a bijection onto $PX'P \in \clb_{\tau}(\clk)$ and the map is norm preserving order isomorphic from the self-adjoint part of the commutant of $\pi_{\psi}(\clo_d)'$ to the space of self-adjoint fixed points of the completely positive map $\tau$. Furthermore, $\clm' = \clb_{\tau}(\clk)$. In particular, for any unitary element $u' \in \clm'$ ( any anti-unitary element $j_0$ commuting with elements in $\clm$ ), there exists a unique unitary element $U'$ in $\pi(\clo)'$ ( an anti-unitary element $\clj_0$ on $\clh$ commuting with elements in $\pi(\clo_d)'' )$ such that 
$PU'P=u'( P \clj_0 P = j_0)$. 

\vsp 
\NI (e) Let $\psi'$ be an another $\lambda$-invariant state of $\clo_d$ with its GNS representation $(\clh_{\psi'},\pi_{\psi'},\zeta_{\psi'})$ and 
$(P',\clk',v'_k:1 \le k \le d)$ be its Popescu system on its support projection $P'=[\pi_{\psi'}(\clo_d)'\zeta_{\psi'}]$ such that 
$$v'_k= u v_k u^*,\;\;1 \le k \le d$$
for some operator $u:\clk \raro \clk'$ then there exists a unique operator $U:\clh_{\psi} \raro \clh_{\psi'}$ extending $u:\clk \raro \clk'$ such that 
$$\pi_{\psi'}(x) = U \pi_{\psi}(x) U^*$$
for all $x \in \clo_d$. Furthermore, the map $U \raro P'UP$ is a bijection between the sets of inter-twinning elements and $U$ is unitary if and only if $u$ is unitary. Similar statement also holds for anti-unitary operators $j_0: \clk \raro \clk'$ and $\clj_0:\clh \raro \clh'$ intertwining the Popescu and the Cuntz elements respectively.   

\vsp
\NI (f) The following statements are equivalent:

\NI (i) $\psi$ is a factor state of $\clo_d$;

\NI (ii) $(\pi_{\psi}(\clo_d)'',\Lambda,\psi)$ is ergodic;

\NI (iii) $\clm$ is a factor;

\NI (iv) $(\clm,\tau,\phi)$ is ergodic.  

\vsp
Conversely, let $v_1,v_2,...,v_d$ be a family of bounded operators on a Hilbert space
$\clk$ so that $\sum_{1 \le k \le d} v_kv_k^*=I$. Then there exists a unique up to
unitary isomorphism Hilbert space $\clh$, a projection operator $P$ on $\clh$ with range equal to $\clk$ 
and a Cuntz element $\{S_k:,\;1 \le k \le d \}$ satisfying relation (10) so that
\be
PS^{*}_{k}P=S_k^*P=v^*_k
\ee
for all $1 \le k \le d$ and $\clk$ is cyclic for the representation i.e. the vectors
$\{ S_I\clk: |I| < \infty \}$ are total in $\clh$.

Moreover, the following holds:

\NI (i) $\Lambda_n(P) \uparrow I$ as $n \uparrow \infty$;

\NI (ii) For any $D \in \clb_{\tau}(\clk)$, $\Lambda_n(D) \raro X'$ weakly as $n \raro \infty$
for some $X'$ in the commutant $\{S_k,S^*_k: 1 \le k \le d \}'$ so that $PX'P=D$. Moreover
the self adjoint elements in the commutant $\{S_k,S^*_k: 1 \le k \le d \}'$ is isometrically
order isomorphic with the self adjoint elements in $\clb_{\tau}(\clk)$ via the
surjective map $X' \raro PX'P$. 

\NI (iii) $\{v_k,v^*_k,\;1 \le k \le d \}' \subseteq \clb_{\tau}(\clk)$ and equality holds,  if and only if
$P \in \{S_k,S_k,\;1 \le k \le d \}''$.

\vsp
\NI (iv) Let $\clm$ be a von-Neumann algebra generated by the family $\{v_k: 1 \le k \le d \}$ of operators on Hilbert space 
$\clk$ and $\clm'=\clb_{\tau}(\clk)$. Then for any  $\tau$-invariant faithful normal state $\phi$ on $\clm$ there exists a 
$\lambda$-invariant state $\psi$ on $\clo_d$, defined by
$$\psi(s_Is^*_J)=\phi(v_Iv^*_J),\;|I|,|J| < \infty $$
so that its GNS space associated with $(\clm,\phi)$ is identified with the support projection of $\psi$ in $\pi_{\psi}(\clo_d)''$, where $(\clh_{\psi},\pi_{\psi},\zeta_{\psi})$ is the GNS space of $(\clo_d,\psi)$. 

\vsp
Furthermore, for a given $\lambda$-invariant state $\psi$, the family $(\clk,\clm,v_k\;1 \le k \le d,\phi)$ satisfying (iv) is determined uniquely up to unitary conjugation. 

\end{pro} 

\begin{proof} 
For (a) we verify the following steps: 
$$\langle \zeta_{\psi}, P\Lambda(I-P)P \zeta_{\psi}\rangle $$
$$=\langle \zeta_{\psi},\Lambda(I-P)\zeta_{\psi}\rangle$$
$$=\langle \zeta_{\psi},(I-P) \zeta_{\psi} \rangle$$
$$=0$$ 
Since $P$ is the support projection of the state $\psi$ on $\pi_{\psi}(\clo_d)''$
and $0 \le P\Lambda(I-P)P \in \pi_{\psi}(\clo_d)''$, we get 
$$P\Lambda(I-P)P=0$$
and so (i) i.e. $(I-P)S_k^*P=0$ for each $1 \le k \le d$ is immediate. 

\vsp 
Since $P\zeta_{\psi}=\zeta_{\psi}$ and $S_J^*P \subseteq P$ for all $|J| < \infty$, we get (ii) by cyclic property of $\zeta_{\psi}$ for $\pi_{\psi}(\clo_d)''$. 

\vsp 
For (iii), let $Y=\mbox{lim}_{n \uparrow \infty}\Lambda^n(P)$
in strong operator limit. Then $Y$ is a projection and $\Lambda(Y)=Y$. By Cuntz relations (17), we get $S_iY=YS_i$. Since $Y^*=Y$, we get $YS_i^*=S_i^*Y$ by taking adjoint on both side. Thus $Y \in \pi_{\psi}(\clo_d)'$. Since $Y \ge P$ and $P\zeta_{\psi}=\zeta_{\psi}$, we get $Y\zeta_{\psi}=\zeta_{\psi}$. Thus 
$$YS_IS_J^*\zeta_{\psi}$$
$$=S_IS_J^*Y\zeta_{\psi}$$
$$=S_IS_J^*\zeta_{\psi}$$
This shows $Y=I$ by cyclic property of $\zeta_{\psi}$ 
for $\pi_{\psi}(\clo_d)''$ in $\clh_{\psi}$.  

\vsp 
We can as well prove (iii) by using the following alternative argument. Since $\Lambda^n(P)S_I=S_IP$ for all $I$ with $|I|=n$, we have by Cuntz relations (17) and (i) that 
$$\Lambda^n(P)S_IS_J^*\zeta_{\psi}$$
$$=S_IPS_J^*\zeta_{\psi}$$
$$=S_IS_J^*\zeta_{\psi}$$
for all $|I|=n$ and $|J| < \infty$. Since $Y \ge \Lambda^n(P)$, we get 
$$YS_IS_J^*\zeta_{\psi}=S_IS_J^*\zeta_{\psi}$$
for all $|I| =n$ and $|J| < \infty$. Since this equality holds for all $n$, we get (iii) by cyclic property of $\zeta_{\psi}$ for $\pi_{\psi}(\clo_d)''$.  

\vsp 
The relation (iv) is a simple computation 
$$\sum_kv_kv_k^*$$
$$=\sum_k PS_kPS_k^*P$$
$$= \sum_k PS_kS_k^*P$$
$$=I_{\clk}$$

\vsp 
The relation (b) follows by (i) of (a). For non trivial statements (c) and (d), we refer to the commutant lifting Theorem 2.1 in [Mo2] as used in Proposition 2.4 in [Mo3]. We also refer Theorem 5.1 in [BJKW] for the original idea used for commutant lifting theorem and [Po]. For a quick recall, we consider the element $U'(U')^* \in \pi(\clo_d)'$ and note that $PU'U'^*P=PU'PPU'^*P=u'u'^*=P$ since $P =[\pi_{\psi}(\clo_d)'\zeta_{\psi}] \in \pi_{\psi}(\clo_d)''$. Thus by the bijective property $U'U'^*=I$. same holds $U'^*U'=I$ as $u'^*u'=P$. In case of an anti-unitary element $j_0$ commuting elements in $\clm$, the weak operator limit of $\Lambda^n(j_0)$ as $n \uparrow \infty$ exists and its limit $\clj_0$ commutes with elements in $\pi(\clo_d)''$. A proof follows a line of argument used in Theorem 2.1 in [Mo2]. That $\clj_0^2=I$ follows as $\clj_0^2$ commutes with elements in $\pi(\clo_d)''$ and $P\clj_0^2P=j^2_0=P$. 
So by the bijective property of the map $X' \raro PX'P$, we get $\clj_0^2=I$ on $\clh$.    

\vsp 
For (e), we consider the representation $\pi \oplus \pi'$ and apply (d) as in Theorem 5.1 in [BJKW] with support projection $P \oplus P'$ in $\clh_{\psi} \oplus \clh_{\psi'}$. 

\vsp 
For (f), we first note that (i) and (ii) are equivalent since the centre of $\pi_{\psi}(\clo_d)''$ are exactly the invariant elements of $\Lambda$ in $\pi_{\psi}(\clo_d)''$. Similarly (iii) and (iv) are equivalent since the invariant elements of $\tau$ in $\clb(\clk)$ i.e. $\clb_{\tau}(\clk)=\clm'$ by (c). Thus the centre of $\clm$ are exactly the invariant elements of $\tau$ in $\clm$. That (ii) and (iv) are equivalent follows 
by the argument used in Theorem 3.6 in [Mo6]. Same method works for discrete time dynamics as well. For a proof, we repeat the argument used now. For any element
$X \in \pi_{\psi}(\clo_d)''$ and any state $\psi'$ on $\pi_{\psi}(\clo_d)''$, we have 
$|\psi'(\Lambda^m(X(I-P)))|^2 \le \psi'(\Lambda^m(X^*X)) \psi'(\Lambda^m(I-P)) \le ||X||^2 \psi'(\Lambda^m(I-P)) \raro 0$ 
as $m \raro \infty$ i.e. $\Lambda^m(X(I-P)) \raro 0$ in weak$^*$ topology of $\pi_{\psi}(\clo_d)''$ as $n \raro \infty$. Thus ${1 \over m} \sum_{0 \le k \le m-1} \Lambda^k(X(I-P)) \raro 0$ in weak$^*$ topology as $m \raro \infty$. 

\vsp 
We also write 
$$\Lambda^{m+n}(PXP)$$
$$= \Lambda^m \Lambda^n(PXP)$$
$$=\Lambda^m(P\Lambda^n(PXP)P) +  \Lambda^m(P^{\perp}\Lambda^n(PXP)P) + \Lambda^m( P\Lambda^n(PXP)P^{\perp}) + \Lambda^m(P^{\perp}\Lambda^n(PXP)P^{\perp})$$
and for any normal state $\psi'$ on $\pi_{\psi}(\clo_d)''$ we note that  
$$\mbox{limsup}_{n \raro \infty} |{1 \over n} \sum_{0 \le k \le n-1} \psi'(\Lambda^{m+k}(PXP))|$$
is independent of $m$, we choose. On the other hand, for $X,Y \in \pi_{\psi}(\clo_d)''$ we have      
$$\mbox{limsup}_{n \raro \infty} |{1 \over n} \psi'(\Lambda^m(\sum_{0 \le k \le n-1} Y\Lambda^k(PXP)P^{\perp})|$$
$$=\mbox{limsup}_{n \raro \infty} |\psi'(\Lambda^m(Y {1 \over n} \sum_{0 \le k \le n-1 }\Lambda^k(PXP)P^{\perp})|$$
$$\le ||Y||\;||X|| \psi'(\Lambda^m(P^{\perp}))^{1 \over 2}$$
Thus $$\mbox{limsup}_{n \raro \infty} |{1 \over n} \psi'(\Lambda^m(\sum_{0 \le k \le n-1} Y\Lambda^k(PXP)P^{\perp})|=0$$

\vsp 
Thus combining the above two steps, we conclude that (ii) and (iv) are equivalent. 

\vsp 
We can prove directly that (i) and (iii) are equivalent as follows. Suppose (i) is true and $a \in \clm \bigcap \clm'$. Then $a = PXP$ for some $X \in \pi_{\psi}(\clo_d)'$ 
since $\tau(a)=a$ and $\clb_{\tau}(\clk)=\clm'$ and $\clb_{\tau}(\clk)=P\pi_{\psi}(\clo_d)'P$ by (d). Thus $\Lambda^n(a)= \Lambda^n(P)X\Lambda^n(P) \in \pi_{\psi}(\clo_d)''$ for all $n \ge 1$. Since $\Lambda^n(P) \uparrow I$, we conclude that $X \in \pi_{\psi}(\clo_d)''$ as 
$X = \mbox{lim}_{n \raro \infty}\Lambda^n(P)X\Lambda^n(P)$ in weak operator topology. Thus $X$ is a scaler multiple of identity operator by the factor property of $\psi$ given 
as (i). So $a = PXP$ is also a scaler multiple of $I_{\clk}$. 

\vsp 
For (iii) implies (i), we take an element $X \in \pi_{\psi}(\clo_d) \bigcap \pi_{\psi}(\clo_d)'$. Then $PXP \in \clm$ since $P\pi_{\psi}(\clo_d)''P=\clm$. However $X \in \pi_{\psi}(\clo_d)'$ and so by (c), we also have $PXP \in \clm'$. So $PXP=\lambda P$ 
for some scaler. Now we use once again action of $\Lambda^n$ on both side and take limit 
$n \raro \infty$ to conclude $X=\lambda I$. Thus (iii) implies (i).    

\end{proof} 

\vsp 
Let $\psi$ be a $\lambda$-invariant state of $\clo_d$ as in Proposition 2.1 and 
$H=\{z \in S^1: \psi = \psi \beta_z \}$ be the closed subgroup of $S^1$. Let $z \raro U_z$ be the unitary representation of $H$ in the GNS space $(\clh_{\psi},\pi,\zeta_{\psi})$ associated with the state $\psi$ of 
$\clo_d$, defined by 
\be 
U_z\pi_{\psi}(x)\zeta_{\psi}=\pi_{\psi}(\beta_z(x))\zeta_{\psi}
\ee
so that $\pi_{\psi}(\beta_z(x))=U_z\pi_{\psi}(x)U_z^*$ for $x \in \clo_d$. We use same notations $(\beta_z:z \in H)$ for its normal extensions as group of automorphisms on $\pi_{\psi}(\clo_d)''$. Furthermore, $\langle \zeta_{\psi},P\beta_z(I-P)P \zeta_{\psi} \rangle=0$ as $\psi=\psi \beta_z$ for $z \in H$. Since $P$ is the support projection of $\psi$ in $\pi_{\psi}(\clo_d)''$, we have $P\beta_z(I-P)P=0$ i.e. $\beta_z(P) \ge P$ for all $z \in H$. Since $H$ is a group, we conclude that $\beta_z(P)=P$ i.e. $PU_z=U_zP$ for all $z \in H$. 

\vsp 
We consider now a group of automorphisms $(\beta_z:z \in H)$ on $\clb(\clk)$ defined by
$\beta_z(a)=u_zau_z^*$, where $z \raro u_z=PU_zP$ is the unitary representation 
of $H$ in $\clk$. Let 
\be
u_z=\sum_{k \in \hat{H}} z^k P_k
\ee 
be 
Stone-Naimark-Ambrose-Godement (SNAG) decomposition [Mac49] of the unitary representation 
$z \raro u_z$ into its dual group $\hat{H}$. So we have 
$$P_k=[v_Iv_J^*\zeta_{\phi}: |I|-|J|=k]$$ for $k \in \hat{H}$. 

\vsp 
Our next two propositions are adapted from results in section 6 and section 7 of [BJKW] as stated in the present form in Proposition 2.5 and Proposition 2.6 in [Mo3]. 

\vsp 
\begin{pro} 
Let $\psi$ be a $\lambda$ invariant factor state on $\clo_d$ and $(\clh_{\psi},\pi_{\psi},\zeta_{\psi})$ be its GNS representation. Then the following holds:

\NI (a) The closed subgroup $H=\{z \in S^1: \psi \beta_z =\psi \}$ is equal to 

$$\{z \in S^1: \beta_z \mbox{extends to an automorphism of } \pi_{\psi}(\clo_d)'' \} $$ 

\NI (b) Let $\clo_d^{H}$ be the fixed point sub-algebra in $\clo_d$ under the gauge group $\{ \beta_z: z \in H \}$. Then  
$\pi_{\psi}(\clo_d^{H})'' = \pi_{\psi}(\mbox{UHF}_d)''$.

\NI (c) If $H$ is a finite cyclic group of $k$-many elements and $\pi_{\psi}(\mbox{UHF}_d)''$ is a factor, 
then $\pi_{\psi}(\clo_d)'' \bigcap \pi_{\psi}(\mbox{UHF}_d)' \equiv \!C^m$,  where $1 \le m \le k$.

\NI (d) If $\pi_{\psi}(\clo_d)''$ is a type-I factor then $H = \{z \in S^1: z^k= 1 \}$, i.e. $H$ is a finite cyclic subgroup of $S^1$ and $\pi_{\psi}(\clo_d)'' \bigcap \pi_{\psi}(\mbox{UHF}_d)' \equiv \!C^k$. If $\pi_{\psi}(UHF_d)''$ is also a factor then $H=\{1\}$ and $\pi_{\psi}(\clo_d)'' = \pi_{\psi}(\mbox{UHF}_d)''$; 

\NI (e) Let $\omega'$ be a $\lambda$-invariant state of $\mbox{UHF}_d$ algebra and  
$\pi_{\omega'}(\mbox{UHF}_d)''$ is a type-I factor, then there exists a $\lambda$-invariant factor state $\psi$ on $\clo_d$ extending $\omega'$ such that 
$$\pi_{\psi}(\mbox{UHF}_d)'' = \pi_{\psi}(\clo_d)''$$
Furthermore, $H= \{ z \in S^1: \psi = \psi \beta_z \}$ is equal to $\{ 1 \}$.    
\end{pro} 

\vsp 
\begin{proof} (a), (b) and (c) are restatement of Proposition 2.5 in [Mo3]. Proofs of (d) and (e) are given in [BJKW] and [Ma3] respectively. Here for our convenience, we give quick proofs as follows. 

\vsp 
If the von-Neumann algebra $\pi_{\psi}(\clo_d)''$ is a type-I factor then the group of 
$*$-automorphism $z \raro \beta_z$ is inner i.e. there exists a unitary representation $z \raro V_z \in \pi_{\psi}(\clo_d)''$ such that 
$$\beta_z(X)=V_zxV_z^*$$
for all $X \in \pi_{\psi}(\clo_d)''$, where $\beta_z(X)=U_zXU^*_z$ for all $X \in \pi_{\psi}(\clo_d)''$ and 
$$U_z \pi_{\psi}(x)\zeta_{\psi}= \pi_{\psi}(\beta_z(x))\zeta_{\psi}$$ 
for all $x \in \clo_d$. For details, we refer to Lemma 6.9 in [BJKW]. 

\vsp 
Let 
\be
V_z=\sum_{k \in \hat{H}} z^k E_k
\ee 
be 
Stone-Naimark-Ambrose-Godement (SNAG) decomposition [Mac49] of 
the unitary representation $z \raro V_z$ into its dual group 
$\hat{H}$. So each $E_k \in \pi_{\psi}(\clo_d)''$. Since $V_z\pi_{\psi}(s_i)V_z^*=z\pi_{\psi}(s_i)$ for all $1 \le i \le d$, we get $\Lambda(V_z)=zV_z$ for all 
$z \in H$, where 
$$\Lambda(X)= \sum_{1 \le k \le d} \pi_{\psi}(s_k)X\pi_{\psi}(s_k^*)$$
for $X \in \pi_{\psi}(\clo_d)''$. By the uniqueness of the decomposition, we get 
$\Lambda(E_k)=E_{k+1}$ for all $k \in \IZ$, if $\hat{H}=\IZ$. Since $\psi \lambda = \psi$, we also have in such a case   
$$\psi(E_{k+1})=\psi(\Lambda(E_k))=\psi(E_k),\; k \in \IZ$$
This brings a contradiction for $\hat{H}=\IZ$ since $1=\psi(I)= \psi ( \sum_{ k \in \IZ} E_k) =\sum_{k \in \IZ} \psi(E_k)$ by normality of $\psi$ on $\pi_{\psi}(\clo_d)''$. Rest of statement of (d) is obvious. 

\vsp    
The last statement (e) uses similar idea that was involved in the proof of (d). Let $\omega$ be the unique inductive limit state of $\IM$ such that $\omega=\omega \theta$ and 
$\omega_{|}\IM_R = \omega'_{|}\mbox{UHF}_d$. We consider the unital injective endomorphism $\Lambda: \pi_{\omega}(\mbox{UHF}_d)'' \raro \pi_{\omega}(\mbox{UHF}_d)''$ defined by extending the map 
$$\Lambda(\pi_{\omega}(x))= \pi_{\omega}(\lambda(x)),\; x  \in \mbox{UHF}_d$$
by restricting on the cyclic space $[\pi_{\omega}(\mbox{UHF}_d)\zeta_{\omega}]$. Since $\pi_{\omega}(\mbox{UHF}_d)''$ is a type-I factor, by a theorem of W. Arveson [Ar] (also see Theorem 3.1 in [BJP]), we get a family of elements $(S_i)$ in $\pi_{\omega}(\mbox{UHF}_d)''$ satisfying Cuntz relation (20) so that 
$$\Lambda(X)= \sum_{1 \le i \le d}S_i X S_i^*$$ 
for all $X \in \pi_{\omega}(\mbox{UHF}_d)'',$
where we verify Arveson index of 
$(\mbox{UHF}_d,\Lambda)$ is $d$ as 
$\mbox{UHF}_d \bigcap \Lambda(\mbox{UHF}_d)'$ 
is isomorphic to $\!M_d(\IC)$. Now we consider the state $\psi:\clo_d \raro \!C$ defined by 
$$\psi(s_Is_J^*) = <\zeta_{\omega}S_IS_J^*\zeta_{\omega}>$$
It is clear that $\psi \lambda = \psi$. We claim that $\psi$ is a factor state. Since 
$[S_IS_J^*\zeta_{\omega}]=\clh_{\omega'}$, we can identify $s_Is_J^* \raro S_IS_J^*$
as a GNS representation of the state $\psi$. Now the factor property of $\psi$ is obvious since by our construction $\pi_{\psi}(\mbox{UHF}_d)''=\pi_{\psi}(\clo_d)''$ as $\pi_{\psi}(s_i)= S_i \in \pi_{\omega}(\mbox{UHF}_d)''$. That $H$ is trivial for $\psi$ now follows by (d). 
\end{proof}

\vsp 
Thus by Proposition 2.2 (b), $P \in \pi_{\psi}(\mbox{UHF}_d)''$ for a $\lambda$-invariant factor state of $\clo_d$. In such a case, we define von-Neumann subalgebra $\clm_0$ of $\clm$ by 
\be 
\clm_0=P\pi_{\psi}(\mbox{UHF}_d)''P
\ee
i.e. $\clm_0$ is weak$^*$ closure of vector space $\{v_Iv_J^*:|I|=|J|\}$ by Proposition 2.1 (a). Let $\clk_0$ be the Hilbert subspace of $\clk$ equal to the range of $[\clm_0\zeta_{\omega}]$. 
Then $\clm_0$ can be realized as a von-Neumann subalgebra of $\clb(\clk_0)$, however 
its commutant in $\clb(\clk_0)$ could be different from commutant $\clm_0'$ taken in $\clb(\clk)$. 

\vsp 
Since endomorphism $\Lambda(X)=\sum_k \pi_{\psi}(s_k)X\pi_{\psi}(s_k^*)$ preserves $\pi(\mbox{UHF}_d)''$, $\tau$ also preserves $\clm_0$. Let $\phi_0$ be the restriction of $\phi$ to $\clm_0$. Thus $(\clm_0,\tau^n,\;n \ge 1, \phi_0)$ is a quantum dynamical system [Mo2] of a completely positive map $\tau$ on $\clm_0$ with a faithful normal invariant state $\phi_0$.    

\vsp
Let $\omega'$ be a $\lambda$-invariant state on the $\mbox{UHF}_d$ sub-algebra of $\clo_d$. Following [BJKW, section 7] and $\omega$ be the inductive limit state $\omega$ of $\IM \equiv \tilde{\mbox{UHF}}_d \otimes \mbox{UHF}_d$. In other words $\omega'=\omega_R$ once we make the identification $\mbox{UHF}_d$ with $\IM_R$.
We consider the set 
$$K_{\omega}= \{ \psi: \psi \mbox{ is a state on } \clo_d \mbox{ such that } \psi \lambda =
\psi \mbox{ and } \psi_{|\mbox{UHF}_d} = \omega_R \}$$
By taking invariant mean on an extension of $\omega_R$ to $\clo_d$, we verify that $K_{\omega}$ is non empty and 
$K_{\omega}$ is clearly convex and compact in the weak topology. In case $\omega$ is an ergodic state ( extremal state ) then, $\omega_R$ is as well an extremal state in the set of $\lambda$-invariant states of $\IM$. Thus
$K_{\omega}$ is a face in the $\lambda$ invariant states. Now we recall Lemma 7.4 
of [BJKW] in the following proposition which quantifies what we can gain 
by considering a factor state on $\clo_d$ instead of its restriction to 
$\mbox{UHF}_d$.

\vsp 
\begin{pro} 
Let $\omega$ be an ergodic state of $\IM$. Then $\psi \in K_{\omega}$ is an extremal point in
$K_{\omega}$,  if and only if $\psi$ is a factor state. Moreover any other extremal point in $K_{\omega}$
is of the form $\psi \beta_z$ for some $z \in S^1$ and 
$H=\{z \in S^1: \psi \beta_z =\psi \}$ is independent of the extremal point $\psi \in K_{\omega}$.   
\end{pro} 

\vsp 
In Proposition 2.1 (b) we have taken an arbitrary element $\psi \in K_{\omega}$ to find 
a Popescu element $\clp=(\clk,v_i \in \clm,1 \le i \le d,\;\zeta_{\omega})$ in its 
support projection and arrived at a representation of $\omega$ given by
\be 
\omega(|e^{i_1}\rangle \langle e_{j_1}|^{(1)} \otimes |e^{i_2}\rangle\langle e_{j_2}|^{(2)} \otimes....\otimes |e^{i_k}\rangle\langle e_{j_k}|^{(k)} \otimes 1 \otimes 1 ..)=\phi(v_Iv_J^*),
\ee
where $I=(i_1,i_2,..,i_k)$ and $J=(j_1,j_2,..,j_k)$. However,  such a representation need 
not be unique even upto unitary conjugation unless $K_{\omega}$ is a singleton set. Nevertheless by Proposition 2.3 for a factor state $\omega$, two extreme points $\psi$ and $\psi'$ in $K_{\omega}$ being related by $\psi'=\psi \beta_z$ for some $z \in S^1$, the Popescu elements $\clp=\{\clk,v_k:1 \le k \le d,\;\sum_kv_kv_k^*=I_{\clk} \}$ 
and $\clp'=\{\clk',v'_k:1 \le k \le d,\; \sum_k v'_k(v'_k)^*=I_{\clk'} \}$ associated with support projections of $\psi$ and $\psi'$ in $\pi_{\psi}(\clo_d)''$ and 
$\pi_{\psi'}(\clo_d)''$ respectively are unitary equivalent modulo a gauge modification i.e. by Proposition 2.1 there exists a unitary operator $u:\clk \raro \clk'$ and $z \in S^1$ so that $uv'_ku^*=z v_k$ for all $1 \le k \le d$. We include more details in the following. We define a unitary operator $U:\clh_{\psi'} \raro \clh_{\psi}$ by extending the inner product preserving map 
$$U \pi_{\psi'}(x) \zeta_{\psi'}= \pi_{\psi}(\beta_z(x))\zeta_{\psi}$$
for all $x \in \clo_d$. So by our construction, we have
$$U \pi_{\psi'}(x)U^*=\pi_{\psi}(\beta_z(x))$$
for all $x \in \clo_d$. In particular, $UP'U^*=P$, where $P=[\pi_{\psi}(\clo_d)'\zeta_{\psi}]$ and $P'=[\pi_{\psi'}(\clo_d)'\zeta_{\psi'}]$. 

\vsp 
We set unitary operator $u:\clk \raro \clk'$ defined by 
$$u=PUP'$$ 
to conclude that 
$$u(v')^*_ku^*$$
$$=PUP'\pi_{\psi'}(s^*_k)P'U^*P$$
$$=PU\pi_{\psi'}(s_k^*)P'U^*P$$
$$=PU\pi_{\psi'}(s_k^*)U^*UP'U^*P$$
$$=PU\pi_{\psi'}(s_k^*)U^*P$$
$$=P\pi_{\psi}(\bar{z}s_k^*))P$$
$$=\bar{z}v^*_k$$
for all $1 \le k \le d$. 

\vsp 
In other words we find a one-one correspondence between 
\be
\omega \Leftrightarrow \omega_R \Leftrightarrow K^{ext}_{\omega} \Leftrightarrow \clp_{ext} \Leftrightarrow (\clm,\tau,\phi) 
\ee  
modulo unitary conjugations and phase factors, where $K^{ext}_{\omega}$ is the set of extreme points in $K_{\omega}$ and $\clp_{ext}$ is the set of Popescu elements associated with extreme points $\psi$ of $K_{\omega}$ on their support projections 
of the states given as in Proposition 2.1. 
Furthermore,  for the support projection $P=[\pi_{\psi}(\clo_d)'\zeta_{\psi}]$, we have $\beta_z(P)=P$ for all $z \in H$. Hence by Proposition 2.2 (b), we have $P \in \pi_{\psi}(\mbox{UHF}_d)''$. So $\clm_0=P\pi_{\psi}(\mbox{UHF}_d)''P$ is a von-Neumann algebra in its own right and $\clm_0 \subseteq \clm$ and $\tau$ takes elements of $\clm_0$ to itself denoted by 
$\tau_0:\clm_o \raro \clm_0$. Thus $(\clm_0,\tau_0^n,\phi_0)$ is a semi-group of unital completely positive maps with a faithful normal invariant state $\phi_0$, the restriction of $\phi$ to $\clm_0$ and such a triplet $(\clm_0,\tau^n_0,\phi_0)$ is canonically associated with the state $\omega$ modulo unitary conjugation. Thus it is natural to expect that various properties of $\omega$ are related to asymptotic properties of $(\clm_0,\tau^n_0,\phi_0)$. We have already explored purity of $\omega$ in [Mo2] to find its precise relation with the asymptotic behaviour of the dynamics $(\clm_0,\tau_0^n,\phi_0)$ as $n \raro \infty$. Along with $(\clm_0,\tau_0^n,\phi_0)$, asymptotic behaviour of the {\it dual dynamics } $(\clm'_0,\tilde{\tau}_0^n,\phi_0)$ ( defined by D Petz [OP] following a work of Accardi-Cecchini [AC] ) also played an important role in our analysis. We now recall the details of it and explain how it is related to symmetry (5) of $\omega$. 

\vsp
Since $\phi$ is a faithful state, $\zeta_{\phi} \in \clk$ is a cyclic and separating vector for $\clm$ and the closure of the closable operator $S_0:a\zeta_{\phi} \raro a^*\zeta_{\phi},\;a \in \clm, S$ possesses a polar decomposition $S=\clj \Delta^{1/2}$, where $\clj$ is an anti-unitary and $\Delta$ is a non-negative self-adjoint operator on $\clk$. M. Tomita [BR] theorem says that $\Delta^{it} \clm \Delta^{-it}=\clm,\;t \in \IR$ and $\clj \clm \clj=\clm'$, where $\clm'$ is the commutant of $\clm$. We define the modular automorphism group
$\sigma=(\sigma_t,\;t \in \IT )$ on $\clm$
by
$$\sigma_t(a)=\Delta^{it}a\Delta^{-it}$$ which satisfies the modular relation
$$\phi(a\sigma_{-{i \over 2}}(b))=\phi(\sigma_{{i \over 2}}(b)a)$$
for any two analytic elements $a,b$ for the group of automorphisms $(\sigma_t)$. A more useful modular relation used frequently in this paper is given by 
\be 
\phi(\sigma_{-{i \over 2}}(a^*)^* \sigma_{-{i \over 2}}(b^*))=\phi(b^*a)
\ee 
which shows that $\clj a\zeta_{\phi}= \sigma_{-{i \over 2}}(a^*)\zeta_{\phi}$ for an analytic element $a$ for the automorphism group $(\sigma_t)$. Anti unitary operator $\clj$ and the group of automorphism $\sigma=(\sigma_t,\;t \in \IR)$ are called {\it conjugate operator} and {\it modular automorphisms } associated with $\phi$ respectively. 

\vsp 
The state $\phi(a)= \langle \zeta_{\phi},x \zeta_{\phi} \rangle $ on $\clm$ being faithful and invariant of $\tau:\clm \raro \clm$, we find a unique unital completely positive map 
$\tilde{\tau}:\clm' \raro \clm'$ ([section 8 in [OP] ) satisfying the duality relation 
\be 
\langle b\zeta_{\phi},\tau(a)\zeta_{\phi} \rangle =  \langle \tilde{\tau}(b)\zeta_{\phi},a\zeta_{\phi} \rangle 
\ee
for all $a \in \clm$ and $b \in \clm'$. For a proof, we refer 
to section 8 in the monograph [OP] or section 2 in [Mo2]. 

\vsp 
Since $\tau(a)=\sum_{1 \le k \le d} v_kav_k^*,\;x \in \clm$ is an {\it inner map } i.e. each $v_k \in \clm$, we have an explicit formula for $\tilde{\tau}$ as follows: For 
each $1 \le k \le d$, we set contractive operator 
\be 
\tilde{v}_k = \overline{ \clj \sigma_{i \over 2}(v^*_k) \clj } \in \clm'
\ee 
That $\tilde{v}_k$ is indeed well defined as an element in $\clm'$ given in section 8 in [BJKW]. By the modular relation (23), we have  
\be  
\sum_k \tilde{v}_k \tilde{v}_k^*=I_{\clk}\;\;\mbox{and}\;\;
\tilde{\tau}(b)=\sum_k \tilde{v}_kb\tilde{v}^*_k,\; b \in \clm' 
\ee
Moreover, if $\tilde{I}=(i_n,..,i_2,i_1)$ for $I=(i_1,i_2,...,i_n)$, we have 
$$\tilde{v}^*_I\zeta_{\phi}$$
$$=\clj \sigma_{i \over 2}(v_{\tilde{I}})^*\clj\zeta_{\phi}$$
$$= \clj \Delta^{1 \over 2}v_{\tilde{I}}\zeta_{\phi}$$
$$=v^*_{\tilde{I}}\zeta_{\phi}$$
and    
\be
\phi(v_Iv^*_J)= \phi(\tilde{v}_{\tilde{I}}\tilde{v}^*_{\tilde{J}}),\; 
|I|,|J| < \infty 
\ee
We also set $\tilde{\clm}$ to be the von-Neumann algebra generated by $\{\tilde{v}_k: 1 \le k \le d \}$. 
Thus $\tilde{\clm} \subseteq \clm'$.

\vsp 
Since $S_0\beta_z(a)\zeta_{\phi}=\beta_z(a^*)\zeta_{\phi}$ for all $a \in \clm$, we have $S_0u_z=u_zS_0$ on $\clm\zeta_{\phi}$. Once again by uniqueness of polar decomposition for $S= \clj \Delta^{1 \over 2}$, we get $u_z \clj u_z^*=\clj$ and $u_z \Delta^{1 \over 2} u_z^*=\Delta^{1 \over 2}$. However, if we write  
$$u_{z} = \sum_{k \in \hat{H}} z^k P_k$$
and then 
$$\clj u_z \clj $$ 
$$=\clj(\sum_{k \in \hat{H}} z^k P_k) \clj$$
$$=\sum_{k \in \hat{H}} \bar{z}^k \clj P_k \clj$$ 
$$=\sum_{k \in \hat{H}} z^k \clj P_{k^{-1}} \clj$$
So $\clj P_k \clj = P_{k^{-1}}$ for all $k \in \hat{H}$. In particular, $\clj$ commutes with $P_k$ if and only if $k = k^{-1}$. 

\vsp 
Furthermore, since $E=\int_{z \in H}\beta_z dz$ is a norm one projection ( i.e. a unital completely positive map $E:\clm \raro \clm_0$ satisfying the bi-module property, i.e. $E(zxy)=zE(x)y,\;x \in \clm, z,y \in \clm_0$ ) from $\clm$ to the fixed point von-Neumann sub-algebra $\clm_0$ of $\clm$, the modular group of automorphisms $(\sigma_t)$ keep $\clm_0$ invariant i.e. $\sigma_t(\clm_0)=\clm_0$ for all $t \in \IR$ by a Theorem of M. Takesaki [Ta]. Note also that $\clm_0=P\pi(\mbox{UHF}_d)''P$ as a von-Neumann algebra with its cyclic space $\clk_0=[\clm_0 \zeta_{\phi}]$. Thus, we have von-Neumann algebra $\clm_0$ acting on $\clk_0$ and the unital completely positive map $\tau_0:a \raro \tau(a),\;a \in \clm_0$ admits a faithful normal invariant state $\phi_0$ on $\clm_0$ which is the restriction of  $\phi$ to $\clm_0$.  Thus $(\clm_0,\tau_0,\phi_0)$ admits an adjoint completely positive map satisfying the duality relation 
given below:
\be 
\langle b\zeta_{\phi},\tau_0(a)\zeta_{\phi} \rangle=\langle \tilde{\tau}_0(b)\zeta_{\phi},a\zeta_{\phi}\rangle
\ee
for all $a \in \clm_0,b \in \clm_0'$, where $\clm_0'$ is the commutant 
of $\clm_0$ in $\clb(\clk_0)$. 

\vsp 
A non-trivial symmetry of $\omega$ will determine a unique affine map on $K_{\omega}$ and thus taking an extremal element of $K_{\omega}$ to another extremal element of $K_{\omega}$. Since associated family of Poposecu elements on support projections of an extremal element are determined uniquely modulo a unitary conjugation, each symmetry will give rises to an undetermined unitary operator intertwining family of Popescu elements modulo a gauge group action. Basic strategy here is to find an algebraic relation between Cuntz state $\psi$ and associated family of Popescu elements $(\clk,v_k:1 \le k \le d)$ in its support projection with its dual Cuntz state $\tilde{\psi}$ associated with dual family of Popescu elements $(\tilde{v}_k,1 \le k \le d)$. While studying symmetry (5) of $\omega$, we need equality of the support projections of $\psi$ and $\tilde{\psi}$ in order to find an algebraic relations between their family of Popescu elements. To that end we recall results from [Mo3] in the next paragraph.      

\vsp 
Let $\tilde{\clo}_d$ be an another copy of Cuntz algebra $\clo_d$ and $\pi$ be the Popescu's prescription [Po] (Theorem 5.1 in [BJKW] or Proposition 2.1 in [Mo3]) 
of a minimal Stinespring representation $\pi: \tilde{\clo}_d \raro \clb(\tilde{\clh})$ associated with the completely positive map $\tilde{s}_I\tilde{s}_J^* \raro \tilde{v}_I\tilde{v}_J^*,\;|I|,|J| < \infty $ so that
$$P\pi(\tilde{s}^*_i)P=\pi(\tilde{s}_i^*)P=\tilde{v}^*_i$$ 
Furthermore, we have 
a {\it dual} state $\tilde{\psi}$ of $\tilde{\clo}_d$, defined by 
$$\tilde{\psi}(\tilde{s}_I \tilde{s}^*_J)=\langle \zeta_{\omega},\tilde{S}_I\tilde{S}^*_J \zeta_{\omega} \rangle$$
\be 
=\phi(\tilde{v}_I\tilde{v}^*_J)
\ee 
However, by the converse part of Proposition 2.1, $P:\tilde{\clh} \raro \clk$ is also the support projection 
of $\tilde{\psi}$ in $\pi(\tilde{\clo}_d)''$, if and only if 
$$ 
\{y \in \clb(\clk): \sum_k \tilde{v}_ky\tilde{v}_k^*= y \} = \tilde{\clm}'
$$

\vsp 
We may define dual $\tilde{\lambda}$-invariant state $\tilde{\psi}$ in the following method as well. 
Let $\psi$ be a $\lambda$-invariant state on $\clo_d$. Let $\tilde{\psi}$ be the state on $\tilde{\clo}_d$, 
which is another copy of $\clo_d$, defined by 
\be 
\tilde{\psi}(\tilde{s}_I\tilde{s}_J^*)=\psi(s_{\tilde{I}}s_{\tilde{J}}^*)
\ee 
for all $|I|,|J| < \infty$. That $\tilde{\psi}$ is well defined and coincide with our earlier definition of 
dual state $\tilde{\psi}$ given in (33) follows once we check by (31) that 
$$\psi(s_{\tilde{I}}s_{\tilde{J}}^*)=\phi(v_{\tilde{I}}v_{\tilde{J}}^*) = 
\phi(\tilde{v}_I\tilde{v}_J^*)$$
for all $|I|,|J| < \infty$. We refer [Mo2], for further details. 

\vsp
Following [BJKW] and [Mo2], we consider the amalgamated tensor product $\clh \otimes_{\clk} \tilde{\clh}$ of $\clh$ with 
$\tilde{\clh}$ over the joint subspace $\clk$. It is the completion of the quotient of the set 
$$\IC \bar{I} \otimes \IC I \otimes \clk,$$ 
where $\bar{I},I$ both consisting of all finite sequences with elements in $\{1,2, ..,d \}$, by the equivalence relation 
defined by a semi-inner product defined on the set by requiring
$$ \langle \bar{I} \otimes I \otimes f,\bar{I}\bar{J} \otimes IJ \otimes g \rangle = \langle f,\tilde{v}_{\bar{J}}v_Jg \rangle, $$
$$ \langle \bar{I}\bar{J} \otimes I \otimes f, \bar{I} \otimes IJ \otimes g \rangle  = \langle \tilde{v}_{\bar{J}}f,v_Jg \rangle $$
and all inner product that are not of these form are zero. We also define two 
commuting representations $(S_i)$ and $(\tilde{S}_i)$ of $\clo_d$ on
$\clh \otimes_{\clk} \tilde{\clh}$ by the following prescription:
$$S_I\lambda(\bar{J} \otimes J \otimes f)=\lambda(\bar{J} \otimes IJ \otimes f),$$
$$\tilde{S}_{\bar{I}}\lambda(\bar{J} \otimes J \otimes f)=\lambda(\bar{J}\bar{I} \otimes J \otimes f),$$
where $\lambda$ is the quotient map from the index set to the Hilbert space. Note that the subspace generated by
$\lambda(\emptyset \otimes I \otimes \clk)$ can be identified with $\clh$ and earlier $S_I$ can be identified
with the restriction of $S_I$ defined here. Same is valid for $\tilde{S}_{\bar{I}}$. The subspace $\clk$ is
identified here with $\lambda(\emptyset \otimes \emptyset \otimes \clk)$. 
Thus $\clk$ is a cyclic subspace for the representation $$\tilde{s}_j \otimes s_i \raro \tilde{S}_j S_i$$ 
of $\tilde{\clo}_d \otimes \clo_d$ in the amalgamated Hilbert space. Let $P$ be the projection on $\clk$. Then we have 
$$S_i^*P=PS_i^*P=v_i^*$$
$$\tilde{S}_i^*P=P\tilde{S}_i^*P=\tilde{v}^*_i$$
for all $1 \le i \le d$. 
We sum up result required in the following proposition.

\begin{pro} 
Let $\psi$ be an extremal element in $K_{\omega}$ and $(\clk,v_k,\;1 \le k \le d)$ be the Popescu elements in the support projection of $\psi$ in $\pi(\clo_d)''$ described in Proposition 2.1 and $(\clk,\tilde{v}_k,\;1 \le k \le d)$ be the dual Popescu elements and $\pi$ be the amalgamated representation of $\tilde{\clo}_d \otimes \clo_d$. Then the following holds:

\NI (a) For any $1 \le i,j \le d$ and $|I|,|J|< \infty$ and $|\bar{I}|,|\bar{J}| < \infty$
$$ \langle \zeta_{\psi},\tilde{S}_{\bar{I}}\tilde{S}^*_{\bar{J}} S_iS_IS^*_JS^*_j \zeta_{\psi} \rangle = \langle \zeta_{\psi}, 
\tilde{S}_i \tilde{S}_{\bar{I}}\tilde{S}^*_{\bar{J}}\tilde{S}^*_jS_IS^*_J \zeta_{\psi} \rangle ;$$

\NI (b) The state $\psi: x \raro \langle \zeta_{\psi},x \zeta_{\psi} \rangle$ defined on  $\tilde{\mbox{UHF}}_d \otimes \mbox{UHF}_d$ is equal to $\omega$ on $\IM$, where we have 
identified   
$$\IM \equiv \IM_{(-\infty, 0]} \otimes \IM_{[1,\infty)} \equiv 
\tilde{\mbox{UHF}}_d \otimes \mbox{UHF}_d;$$ 
with respect to an orthonormal basis $e=(e_i)$ of $\IC^d$. 

\NI (c) $\pi(\tilde{\clo}_d \otimes \clo_d)''= \clb(\tilde{\clh} \otimes_{\clk} \clh)$ and 
$\clm \vee \tilde{\clm} = \clb(\clk)$;

\NI (d) If $\omega$ is a factor state then $\pi(\clo^{H}_d))''=\pi(\mbox{UHF}_d)''$ and
$\pi(\tilde{\clo}^{H}_d)''=\pi(\mbox{UHF}_d)''$;

\NI (e) Let $E$ and $\tilde{E}$ be the support projection of vector state given by $\zeta_{\psi}$ in $\pi(\clo_d)''$ and $\pi(\tilde{\clo})d)''$ respectively i.e. $E=[\pi(\clo_d)'\zeta_{\psi}]$ and $\tilde{E}=[\pi(\tilde{\clo}_d)'\zeta_{\psi}]$. If $F=[\pi(\clo_d)''\zeta_{\psi}]$ and $\tilde{F}=[\pi(\tilde{\clo}_d)''\zeta_{\psi}]$ then, the following statements are equivalent:

\NI (i) $\omega$ is pure;

\NI (ii) $E=\tilde{F}$ and $\tilde{E}=F$; 

\NI (iii) $P=\tilde{E}F$;

\NI (iv) $\tilde{\clm}'=\clb(\clk)_{\tilde{\tau}}$; 

\NI (v) $\pi_{\omega}(\IM_R)'=\pi_{\omega}(\IM_L)''$ (Haag duality) in the 
GNS space $(\clh_{\omega},\pi_{\omega},\zeta_{\omega})$ associated with the state 
$\omega$ of $\IM$.

\vsp 
In such a case ( i.e. if $\omega$ is pure ) $\clm'=\tilde{\clm}$ as von Neumann 
sub-algebra of $\clb(\clk)$, $\clm'_0= \tilde{\clm}_0$ as von-Neumann sub-algebra of $\clb(\clk_0)$ and $\clm_0 \vee \tilde{\clm}_0 = \clb(\clk_0)$. Furthermore, $\tilde{\clm}=P\pi(\tilde{\clo}_d)''P$ and $\tilde{\clm}_0=P\pi(\tilde{\mbox{UHF}}_d)''P$. 
\end{pro} 

\vsp 
\begin{proof} 
For (a)-(d) we refer to Proposition 3.1 and Proposition 3.2 in [Mo3]. For (e) we refer to Theorem 3.6 in [Mo3]. 
\end{proof}

\vsp 
\begin{rem} 
Let $\omega$ be a translation invariant factor state of $\IM$ and $\psi$ be an extremal element in $K_{\omega}$. Thus 
$\psi$ is a factor state and $$\pi_{\psi}(\mbox{UHF}_d)'' = \{ \beta_z(X)=X; X \in \pi_{\psi}(\clo_d)'' \forall z \in H \}$$ by Proposition 2.2 (b). The support projection $P=[\pi_{\psi}(\clo_d)'\zeta_{\psi}$ of the state $\psi$ in an element in $\pi_{\psi}(\clo_d)''$ satisfies $\beta_z(P)=P$ for all $z \in H \}$. The group of automorphism $\beta_z:\pi_{\psi}(\clo)'' \raro \pi_{\psi}(\clo_d)''$ has a natural restriction on $\clm= P\pi_{\psi}(\clo_d)''P$, defined by 
$\beta_z:a \raro \beta_z(PaP)$ for all $x \in \clm$. In particular, we have $\beta_z(a)=a$ for all $a \in \clm_0$, 
where $\clm_0$ is a von-Neumann sub-algebra of $\clm$ defined by $\clm_0=P\pi_{\psi}(\mbox{UHF}_d)''P$. It is simple 
to show 
$$\clm_0= \{ x \in \clm: \beta_z(x),\forall z \in H \}$$ 
Suppose $\beta_z(a)=a$ for some $a \in \clm$ then $\beta_z(PaP)=\beta_z(P)\beta_z(a)\beta_z(P)=PaP$ and $PaP \in \pi_{\psi}(\clo_d)''$. Thus $PaP \in \pi_{\psi}(\mbox{UHF}_d)''$. Since $a=P(PaP)P$, we conclude that $a \in \clm_0$.    
\end{rem}  
 
\vsp 
\begin{rem}
Let $\omega$ be also pure. Then $P=E\tilde{E}$ and $F_0\pi_{\omega}(\tilde{\mbox{UHF}}_d \mbox{UHF}_d)''F_0$ is the algebra of all bounded operators on the closed subspace $F_0= [\pi_{\omega}(\tilde{\mbox{UHF}}_d \mbox{UHF}_d)''\zeta_{\psi}]$. Since $E \in \pi_{\psi}(\clo_d)''$ and $\tilde{E} \in \psi_{\psi}(\tilde{\clo}_d)''$, 
we can verify the following equalities: 
$$P\pi_{\omega}(\tilde{s}_{I'}\tilde{s}_{J'}^* \otimes s_Is^*_J)\zeta_{\psi}$$
$$=\tilde{E}E(\tilde{S}_{I'}\tilde{S}_{J'}^* S_IS^*_J) \tilde{E}E \zeta_{\psi}$$
$$=E\tilde{E}(\tilde{S}_{I'}\tilde{S}_{J'}^*E\tilde{E} S_IS^*_J) \tilde{E}E \zeta_{\psi}$$
$$=\tilde{v}_{I'}\tilde{v}_J^*v_Iv_J^*\zeta_{\psi}$$
for all $|I'|,|J'|,|I|$ and $|J| < \infty$. In particular, we get 
$PF_0=P_0$ since $P_0=\{ f \in \clk: u_zf=f,\;\forall z \in H \}$ and $P_0=[\clm_0\zeta_{\psi}]$. 
Thus, we have $P_0=PF_0=F_0P$. The von-Neumann algebra $P_0\tilde{\clm}_0 \vee \clm P_0$ is equal to 
the algebra of all bounded operators on $P_0$. Thus both $P_0 \clm P_0$ and $P_0\tilde{\clm}P_0$ are factors acting 
on $P_0$. Let $x$ be an element in the centre of $\clm_0$. Then $x$ commutes with $P_0$ and all operators in $\clm_0 \vee \tilde{\clm}$. In particular, $x$ commutes with all operators in $P_0 \clm_0 \vee \tilde{\clm}_0P_0$. Thus pure 
property of $\omega$ ensures $xP_0 = \lambda P_0$ for some scaler $\lambda \in \IC$. Since elements in 
$\tilde{\clm}$ commutes with $x$ and $[\tilde{\clm}\zeta_{\psi}]=\clk$, we conclude $x =\lambda I_{\clk}$. This shows that $\clm_0$ is a factor if $\omega$ is a pure state of $\IM$. 

\vsp 
Let $X$ be an element in the centre of $\pi(\mbox{UHF}_d)''$. Then $X \in \pi(\tilde{\mbox{UHF}}_d \otimes \mbox{UHF}_d)'$ and $X$ commutes with $\{U_z:z \in H \}$. The state $\omega$ being pure, $\{U_z:z \in H \}'= \{U_z: z \in H \} \vee \pi(\tilde{\mbox{UHF}}_d \otimes \mbox{UHF}_d)''$ by Proposition 3.4 (b) in [Mo2]. This shows $X \in \{U_z: z \in H \}''$ and thus $X = \sum_k c_k F_k$. where $U_z= \sum_{k \in \hat{H}} z^k F_k$ is the SNAG decomposition in $\tilde{\clh} \otimes_{\clk} \clh$ as described in Proposition 3.4 in [Mo2]. But $X$ also commutes with elements in $\pi(\tilde{\clo}_d)''$ and $\tilde{\Lambda}(F_k)=F_{k+1}$ and so $c_k=c$ for all $k \in \hat{H}$ and thus $X=cI$. We conclude 
that $\pi(\mbox{UHF})''$ is also a factor if $\omega$ is pure. 
\end{rem}

\vsp
Let $G$ be a compact group and $g \raro u(g)$ be a $d-$dimensional unitary representation of $G$. By $\gamma_g$, we denote
the product action of $G$ on the infinite tensor product $\IM$ induced by $u(g)$,
$$\gamma_g(Q)=(..\otimes u(g) \otimes u(g)\otimes u(g)...)Q(...\otimes u(g)^*\otimes u(g)^*\otimes u(g)^*...)$$
for any $Q \in \IM$. We recall now that the canonical action of the group $U_d(\IC)$ of $d \times d$ matrices on
$\clo_d$ is given by
$$\beta_{u(g)}(s_j)=\sum_{1 \le i \le d} s_i u(g)^i_j $$
and thus
$$\beta_{u(g)}(s^*_j) = \sum_{1 \le i \le d} \bar{u(g)^i_j} s^*_i $$

\vsp
Note that $u(g)|e_i><e_j|u(g)^*=|u(g)e_i><u(g)e_j| = \sum_{k,l} u(g)^l_i \bar{u(g)}^k_j|e_l><e_k|$, where $e_1,..,e_d$
are the standard basis for $\IC^d$. Identifying $|e_i><e_j|$ with $s_is^*_j$, we verify that on $\IM_R$ the gauge action
$\beta_{u(g)}$ of the Cuntz algebra $\clo_d$ and $\gamma_g$ coincide i.e. $\gamma_g(Q)=\beta_{u(g)}(Q)$
for all $Q \in \IM_R$.

\vsp
\begin{pro} 
Let $\omega$ be a translation invariant factor state on $\IM$. Suppose that
$\omega$ is $G-$invariant,
$$\omega(\gamma_g(Q))=\omega(Q) \; \mbox{for all } g \in G \mbox{ and any } Q \in \IM. $$
Let $\psi$ be an extremal point in $K_{\omega}$ and $(\clk,\clm,v_k,\;1 \le k \le d,\phi)$ be the Popescu system associated with $(\clh, S_i=\pi(s_i),\zeta_{\psi})$, described as in Proposition 2.4. Then we have the following:

\NI (a) There exists a unitary representation $g \raro \hat{U}(g)$ in $\clb(\tilde{\clh} \otimes_{\clk} \clh)$ and a representation $g \raro \zeta(g) \in S^1$ 
so that
\be
\hat{U}(g)S_i\hat{U}(g)^*= \zeta(g) \beta_{u(g)}(S_i),\;1 \le i \le d
\ee
and
\be 
\hat{U}(g)\tilde{S}_i\hat{U}(g)^*= \zeta(g) \beta_{u(g)}(\tilde{S}_i),\;1 \le i \le d
\ee
for all $g \in G$. 

\NI (b) There exists a unitary representation $g \raro \hat{u}(g)$ in $\clb(\clk)$ so that $\hat{u}(g)\clm \hat{u}(g)^*=\clm$ for
all $g \in G$ and $\phi(\hat{u}(g)x\hat{u}(g)^*)=\phi(x)$ for all $x \in \clm$. Furthermore, the operator
$V^*=(v^*_1,..,v^*_d)^{tr}: \clk \raro \IC^d \otimes \clk $ is an isometry which intertwines the representation
of $G$,
\be
( \zeta(g) \hat{u}(g) \otimes u(g) )V^*=V^*\hat{u}(g)
\ee
for all $g \in G$, where $g \raro \zeta(g)$ is the representation of $G$ in $U(1)$.

\NI (c) $\clj \hat{u}(g) \clj=\hat{u}(g)$ and $\Delta^{it}\hat{u}(g)\Delta^{-it}=\hat{u}(g)$ for all $g \in G$ and $t \in \IR$.

\NI (d) $u_zu(g) = u(g)u_z$ for all $g \in G$ and $z \in H$. 

\end{pro}

\vsp
\begin{proof} We recall from (18) that 
$$\lambda \beta_g = \beta_g \lambda$$ 
for all $g \in G$ and $\omega$ being $G$-invariant, we have $\psi \beta_g \in K_{\omega}$ for all $\psi \in K_{\omega}$ 
and $g \in G$ and the map $\psi \raro \psi \beta_g$ is an affine one to one and onto map 
on $K_{\omega}$. Thus $\psi \beta_g$ is an extremal element in $K_{\omega}$ if and only if $\psi$ is so. 

\vsp 
Now we fix an extremal element in $K_{\omega}$. The state $\omega$ being a factor state, by Proposition 2.3, any other extremal element $\psi' \in K_{\omega}$ is determined by $\psi' = \psi \beta_z$ for some $z \in S^1$.  

\vsp
The subgroup $H=\{ z \in S^1: \psi = \psi \beta_z \}$ of $S^1$ being close, $H$ is either $S^1$ or a finite cyclic subgroup. In case $H=S^1$, by Proposition 2.3, $K_{\omega}$ is having a unique element and thus by our starting remark we have $\psi \beta_g = \psi$ for all $g \in G$ and $\psi$ is the unique extremal element in $K_{\omega}$. In such a case, we define unitary operator on $\tilde{\clh} \otimes_{\clk} \clh$ by 
$$\hat{U}(g)\pi(\tilde{s}_{I'}\tilde{s}^*_{J'} \otimes s_Is^*_J)\zeta_{\psi} 
= \pi(\beta_{u(g)}(\tilde{s}_{I'}\tilde{s}^*_{J'} \otimes s_Is^*_J))\zeta_{\psi}$$
for all $|I'|, |J'|, |I|,|J| < \infty$ and verify (a) with 
$\zeta(g)=1$ for all $g \in G$.

\vsp
Now we are left to deal with the more delicate case. Let $H = \{z: z^n=1 \}$ for some $n \ge 1$. We define a unitary representation $z \raro U_z$ of $H$ on $\tilde{\clh} \otimes_{\clk} \clh$ as in Proposition 2.3 in [Mo3] by 
$$U_z \pi(\tilde{s}_{I'}\tilde{s}^*_{J'} \otimes s_{I}s_{J}^*)\zeta_{\psi} = z^{|I'|+|I|-|J|-|J'|}
\pi(\tilde{s}_{I'}\tilde{s}^*_{J'} \otimes s_{I}s_{J}^*)\zeta_{\psi}$$
and write its spectral representation  
$$U_z = \sum_{1 \le k \le n-1} z^k F_{k}$$ 
The mutually orthogonal family $\{ F_k: 1 \le k \le n-1 \}$ of projections satisfies in particular  
$$\sum_{0 \le k \le n-1} F_k= I$$ 

\vsp 
By Proposition 2.4 (d), we have $\pi(\clo^H_d)''=\pi(\mbox{UHF}_d)''$ and $\pi(\tilde{\clo}^H_d)''=\pi(\tilde{\mbox{UHF}}_d)''$. Thus for each $0 \le k \le n-1$, 
the orthogonal projection $F_k$ is spanned by the vectors 
$$\{\tilde{S}_{I'}\tilde{S}^*_{J'}S_IS^*_JS^*_K\zeta_{\psi}: |I'|=|J'|,|I|=|J|,\; \mbox{and}\; |K|=k \}$$

\vsp 
We set unitary operator $U'(g)$ on $F_k: k \ge 0$ by 
$$U'(g)\pi(\tilde{s}_{I'}\tilde{s}^*_{J'}s_Is^*_Js^*_K)\zeta_{\psi} =
\pi(\beta_{u(g)}(\tilde{s}_{I'}\tilde{s}^*_{J'} \otimes s_Is^*_Js^*_K))\zeta_{\psi}$$
where $|I'|=|J'|,|I|=|J|$ and $|K|=k$. It is a routine work to check that $U'(g)$ 
is indeed an inner product preserving map on the total vectors in $F_k$ using 
our assumption that $\omega = \omega \beta_g$ on $\IM$. Thus each $U'(g)$ extends 
uniquely to a unitary operator on $\tilde{\clh} \otimes_{\clk} \clh$ and $g \raro U'(g)$ 
is a representation of $G$ in $\clh \otimes_{\clk} \tilde{\clh}$.

\vsp
For each $g \in G$ the Popescu element $(\clk,\beta_{u(g)}(v_k),1 \le k \le d,\;\zeta_{\psi})$ determines an extremal point
$\psi_g \in K_{\omega}$ and thus by Proposition 2.3 there exists a complex number $\zeta(g)$ with modulus $1$
so that $\psi_{g}=\psi \beta_{\zeta(g)}$. Note that for another such a choice $\zeta'(g)$, we have $\bar{\zeta(g)}\zeta'(g) \in H$.
As $H$ is a finite cyclic subgroup of $S^1$, we have a unique  choice once we take $\zeta(g)$ to be an element in the group
$S^1 / H$ which we identify with $S^1$. That $g \raro \zeta(g)$ is a representation of $G$ in $S^1=\{z \in \IC: |z|=1 \}$
follows as the choice in $S^1 / H$ of $\zeta(g)$ is unique. For each $g \in G$, we define a unitary operator by 
$$\hat{U}(g) = \sum_{0 \le k \le n-1} \zeta(g)^k U'(g)F_k$$ 
Both $g \raro U'(g)$ and $g \raro \zeta(g)$ being representations of $G$, 
we conclude that $g \raro U(g)$ is a unitary representation of $G$. So by our construction, we have 
$$\hat{U}(g)\zeta_{\psi}=\zeta_{\psi}$$
$$\hat{U}(g)S_i\hat{U}(g)^* = \zeta(g)\beta_{u(g)}(S_i)$$
and
$$\hat{U}(g)\tilde{S}_i\hat{U}(g)^* = \zeta(g)\beta_{u(g)}(\tilde{S}_i)$$
for all $1 \le i \le d$. 

\vsp
By covariance relations (39), $\hat{U}(g)\pi(\clo_d)''\hat{U}(g)^* = \pi(\clo_d)''$ 
and so $\hat{U}(g)\pi(\clo_d)'\hat{U}(g)^*=\pi(\clo_d)'$ for all $g \in G$. 
So $\hat{U}(g) F \hat{U}(g)^*=F$ and $\hat{U}(g)E \hat{U}(g)^*=E$, where we recall 
$F=[\pi(\clo_d)''\zeta_{\psi}]$ and $E=[\pi(\clo_d)'\zeta_{\psi}]$. 
Hence the support projection $P=E F$ of the state $\psi$ in $\pi(\clo_d)''F$ is also $G$ 
invariant i.e. $\hat{U}(g)P\hat{U}(g)^*=P$ for all $g \in G$. Thus we define 
a unitary representation of $g$ in $\clk$ by 
$g \raro \hat{u}(g)=P\hat{U}(g)P,\; g \in G$. 
Hence we have 
$$\hat{u}(g)v_j\hat{u}(g)^*=\zeta(g) \beta_{u(g)}(v_j) = \zeta(g)v_i u(g)^i_j$$
for all $1 \le j \le d$, where we recall $v_j = PS_jP$. By taking adjoint of the equation above, we get $$\hat{u}(g)v^*_j\hat{u}(g)^*=\bar{\zeta(g)}\bar{u(g)^i_j} v^*_i$$ 
for all $1 \le j \le d$.

\vsp
We are now left to prove (c). Since automorphism $a \raro \hat{u}(g)a\hat{u}(g)^*$ on $\clm$ preserves the faithful normal state $\phi$, it follows from 
a general theorem due to A. Frigerio [Fr], see also [OP] that the modular group and conjugate operator associated with $\phi$ commutes with the auto-morphism. Proof uses the fact that KMS property uniquely determines modular group by a theorem of M. Takesaki [Ta2].  
Otherwise also we can verify directly here that the densely defined closable Tomita conjugation operator $S_0x\zeta_{\psi}=x^*\zeta_{\psi}$ for $x \in \clm$ 
satisfies $S_0\hat{u}(g)=\hat{u}(g)S_0$ as their actions on any typical vector $v_Iv^*_J\zeta_{\psi}$ are same by the covariance relation (41), where $S_0x\zeta_{\psi}=x^*\zeta_{\psi}$ for $x \in \clm$. Hence by the uniqueness of the polar decomposition, we conclude that (c) holds. 
\end{proof}

\section{Translation invariant lattice refection symmetry state with a twist $r_0$ }

\vsp 
For a given $u \in U_d(\IC)$, we extend the map $\tilde{\beta}_u:\IM \raro \IM$ defined in (7) to an automorphism on 
$\tilde{\clo}_d \otimes \clo_d$, defined by 
\be 
\tilde{\beta}_u(\tilde{s}_{I'}\tilde{s}^*_{J'} \otimes s_Is_J^*) = \beta_{u}(\tilde{s}_I\tilde{s}_J \otimes s_{I'}s^*_{J'})
\ee 
for all $|I|,|J|,|I'|,|J'| < \infty$ and then extend linearly for an arbitrary element of $\tilde{\clo}_d \otimes \clo_d$. So we have 
\be 
\tilde{\beta}_u=\beta_u \tilde{\beta}_{_{I_d}} = \tilde{\beta}_{_{I_d}} \beta_u
\ee

\vsp 
For a given $u \in U_d(\IC)$, we also extend the map $\clj_u:\IM \raro \IM$ defined in (10) to an anti-linear automorphism on 
$\tilde{\clo}_d \otimes \clo_d$, defined by 
\be 
\clj_u(\tilde{s}_{I'}\tilde{s}^*_{J'} \otimes s_Is_J^*) = \beta_{\bar{u}}(\tilde{s}_I\tilde{s}_J \otimes s_{I'}s^*_{J'})
\ee 
for all $|I|,|J|,|I'|,|J'| < \infty$ and then extend anti-linearly for an arbitrary element of 
$\tilde{\clo}_d \otimes \clo_d$. So we have 
\be 
\clj_u = \beta_{\bar{u}} \clj_{_{I_d}}= \clj_{_{I_d}} \beta_{u}
\ee 

\vsp 
So these maps are defined after fixing the orthonormal basis $e=(e_i)$, which have identified $\tilde{\mbox{UHF}} \otimes \mbox{UHF}_d$ with $\IM_L \otimes \IM_R= M$ as in Proposition 2.4 (b), where
the monomial given (21) is identified with the matrix given in (22).  

\vsp 
For $u,w \in U_d(\IC)$, we have 
$$\tilde{\beta}_{u}\tilde{\beta}_{w}$$
\be 
=\beta_{uw} 
\ee
and so 
$$\tilde{\beta}_{w}\tilde{\beta}_{u}$$
$$=\beta_{wu}$$
$$=\beta_{\bar{u}w}$$
(provided $wu=\bar{u}w$)
\be 
=\tilde{\beta}_{\bar{u}} \tilde{\beta}_{w}
\ee

\vsp  
We make few simple observations in the following for $u,w \in U_d(\IC)$:
$$\clj_{u}\clj_{w}$$
$$=\beta_{\bar{u}}\clj_{_{I_d}} \clj_{_{I_d}} \beta_w$$ 
\be 
= \beta_{\bar{u}w},
\ee  
and 
$$\clj_{w}\beta_{u}$$
$$=\clj_{_{I_d}} \beta_w \beta_u$$
$$=\clj_{_{I_d}} \beta_{wu}$$
\be 
=\clj_{wu}
\ee 

Also
$$\beta_u \clj_{w}$$
$$=\beta_u \clj_{_{I_d}} \beta_w$$
$$=\clj_{_{I_d}}\beta_{\bar{u}}\beta_w$$
$$=\clj_{_{I_d}}\beta_{\bar{u}w}$$
\be 
=\clj_{\bar{u}w}
\ee
i.e. $\clj_w$ commutes with $\beta_u$ if $wuw^*=\bar{u}$. 

\vsp
By combining relations (52)-(53), we have the following identities
$$\clj_w\beta_{u}$$
$$=\clj_{wu}\; \mbox{by}\; (52)$$
$$=\clj_{\bar{u}w}\; (\mbox{provided}\; wuw^*=\bar{u} )$$
\be 
=\beta_u\clj_w\;\mbox{by}\;(53)
\ee

\vsp 
Let $G$ be the simply connected Lie group $SU_2(\IC)$ and $g \raro u(g)$ be a $d-$dimensional unitary irreducible representation of $G$. Then there exists a $r \in U_d(\IC)$ such that 
\be 
r u(g) r^*=\bar{u(g)}
\ee  
for all $g \in G$. The element $r \in U_d(\IC)$ is determined uniquely modulo a phase factor in $S^1$. In particular any element
\be 
r_z=zr_0,\;z \in S^1
\ee 
satisfies (55), where we have fixed a $r_0$ satisfying (55) with additional condition 
\be 
r_0^2=I_d
\ee
In our notions $r_1=r_0$. In such a case, $r_{-1}=-r_0$ is the only other choice that satisfies (55) and (57) instead of $r_0$. 

\vsp 
With such a choice for $r_0$, we have 
$$\bar{r}_0r_0 u(g)r^*_0\bar{r}_0^*$$
$$=\bar{r}_0\bar{u(g)}\bar{r}_0^*$$
$$=\overline{r_0u(g)r_0^*}$$
$$=\overline{\overline{u(g)}}$$
$$=u(g)$$ 
for all $g \in G$. So by the irreducible property of the representation $g \raro u(g)$ and commuting property $r_0\bar{r}_0=\bar{r}_0r_0$, we conclude that $\bar{r}_0r_0 = \mu I_d$, where $\mu$ is a real number of modulus one. Thus $\bar{r}_0=\mu r_0$ as $r_0^2=I_d$. Taking determinants of matrices on 
both sides of $\bar{r}_0r_0=\mu I_d$, we get $\mu^d=det(r_0)det(\bar{r}_0) = |det(r_0)|^2 =1$. This 
shows that $\mu=1$ if $d$ is an odd integer. For even values of $d$, we make a direct calculation to show $\mu=-1$ as follows: 

\vsp 
For $d=2$, let $\sigma_x,\sigma_y$ and $\sigma_z$ be the Pauli matrices
in $\!M_2(\IC)$ (see the last part of section 4). The self-adjoint matrix $\sigma_y$ is also a  
unitary  matrix i.e. $\sigma_y^2=I_2$ and 
$$\sigma_y \sigma_x \sigma_y = - \sigma_x$$ 
and 
$$\sigma_y \sigma_z \sigma_y = -\sigma_z$$
Since $\sigma_x = \bar{\sigma_x}$ and $\sigma_z = \bar{\sigma_z}$, 
$\sigma_y$ inter-twins $e^{it\sigma_x}$ and $e^{it\sigma_z}$ with their conjugate matrices $e^{-it\sigma_x}$ and $e^{-it\sigma_z}$ respectively for all $t \in \IR$. In contrast, since $\bar{\sigma_y}=-\sigma_y$, we also get $\sigma_y$ inter-twins $e^{it\sigma_y}$ with $e^{-it\sigma_y}$ for all $t \in \IR$. So we set $r_0=\sigma_y$ ( other choice we can make for $r_0$ is $-\sigma_y$) and verify directly that $\bar{r_0}=-r_0$ i.e. $\mu=-1$ if $d=2$. 

\vsp 
We write $i\sigma_y=e^{it_0\sigma_y} \in SU_2(\IC)$, where $t_0={\pi \over 2}$ and verify that
$$u(e^{it_0\sigma_y})u(g)u(e^{-it_0\sigma_y})$$
$$=u(i\sigma_y) u(g) u(i\sigma_y)^*$$
$$=u((i\sigma_y) g (i\sigma_y)^*)$$
$$=u(\bar{g})$$
Since $su_2(\IC)$ is a real Lie algebra that has unique Lie algebra extension to a complex Lie algebra $sl_2(\IC)$, 
i.e. Lie algebra over the field of complex numbers, we also have 
$$u(\bar{g})=\bar{u(g)}$$ 
for all $g \in SU_2(\IC)$ ( Lie-derivatives of the representations in both sides are equal as element in $sl_2(\IC)$). 
So we have  
\be 
u(e^{it_0\sigma_y})u(g)u(e^{-it_0\sigma_y})
=\overline{u(g)}
\ee
If $\pi_u$ is the associated Lie-representation of $su_2(\IC)$, we have 
$$u(e^{it_0\sigma_y})=e^{it_0\pi_{u}(\sigma_y)}$$
for even integer values of $d$, whereas 
$$u(e^{it_0\sigma_y})=e^{2it_0\pi_{u}(\sigma_y)}$$ 
for odd integer values of $d$. Thus for an arbitrary even values of $d$, the unitary matrix $r_0 = e^{it_0\pi_{u}(\sigma_y)})$ satisfies (55) and (57). In contrast, for an arbitrary odd values of $d$, the unitary matrix $r_0=e^{i2t_0\pi_u(\sigma_y)}$ satisfies (55) and (57). In short, $\mu=1$ if $d$ is an odd integer and $-1$ if $d$ is an even integer. 

\vsp 
We write $\mu=\zeta^2$ and set $r_0 \in U_d(\IC)$, such that 
$$\zeta r_0 = u(e^{it_0\sigma_y}) \in U_d(\IC),$$  
where $\zeta^2=\mu$ and so $\mu$ is $1$ for odd values of $d$ otherwise $-1$. 
In the last section, we will recall standard explicit description of $r_0$ and $g \raro u(g)$ 
that satisfies (55) and (57). Note also that $r_{\zeta}=\zeta r_0$ is a matrix with real entries 
irrespective of values taken for $d$. 

\vsp 
The irreducible property of the representation $g \raro u(g)$ is only used to ensure existence of a family of 
$\{r_z \in U_d(\IC) \}$ satisfying (55) and (57). But the irreducibility property is not necessary for 
a more general situation. As an example, same relations are valid if we consider the representation $g \raro u(g) \otimes u(g) \otimes ..\otimes u(g)$ in a finite or infinite tensor product representation of an irreducible one $g \raro u(g)$. This observation is useful, when we investigate the present problem in a quasi one dimensional lattice with   
$$\IM(n) = \otimes_{\ul{j} \in \IZ \times \IZ_n } \IM_d^{(\ul{j})}$$ 
with $\IZ_n= \{m :0 \le m \le n-1 \}$ or a higher dimensional latices say on  
$$\IM_k = \otimes_{\ul{j} \in \IZ_k} \IM_d^{(\ul{j})}$$ 
where $\IZ_k = \IZ \times \IZ ..\times \IZ$ is the $k$ dimensional lattice of integers.  

\vsp 
Now we go back to our main text. So we have 
\be 
\clj_{r_z}^2(x)=\beta_{\bar{r_z}r_z}(x)=\beta_{\mu I_d}(x)
\ee 
for all $x \in \tilde{\clo}_d \otimes \clo_d$, where $\mu=1$ or $-1$ depending on $d$ odd or even. In any case, by 
(54) and (55), we also have  
\be 
\clj_{r_z} \beta_{u(g)} = \beta_{u(g)} \clj_{r_z}
\ee
for all $g \in SU_2(\IC)$.  

\vsp 
Let $\omega$ be a translation invariant factor state of $\IM$ and $\psi$ be an extremal element in $K_{\omega}$. We define a state $\psi_0: \tilde{\clo}_d \otimes \clo_d \raro \IC$ by extending both $\tilde{\psi}: \tilde{\clo}_d \raro \IC$ and $\psi:\clo_d \raro \IC$ by 
\be 
\psi_0(\tilde{s}_{I'}\tilde{s}^*_{J'} \otimes s_Is_J^*) 
= < \zeta_{\psi}, \tilde{v}_{I'}\tilde{v}^*_{J'}v_I^*v_J^* \zeta_{\psi}>
\ee
for all $|I'|,|J'|,|I|$ and $|J| < \infty$. Proposition 2.4 says that $(\tilde{\clh} \otimes_{\clk} \clh, \pi, \zeta_{\psi})$ is the GNS representation $(\clh_{\psi_0},\pi_{\psi_0},\zeta_{\psi_0})$ of $(\tilde{\clo}_d \otimes \clo_d,\psi_0)$. 
For details, we refer to [Mo2]. 

\vsp 
Let $\omega$ be also $SU_2(\IC)$ invariant with a unitary representation (need not be irreducible) $g \raro u(g)$ on $\IC^d$ satisfying (13). Let $\psi$ be an extremal element in $K_{\omega}$. Then by Proposition 2.3, $\psi \beta_{u(g)} = \psi \beta_{\chi(g)I_d}$ on $\clo_d$ for some unique $\chi(g) \in S^1 / H$. Since $\beta_{u(g)}\beta_{u(h)}=\beta_{u(gh)}$, the map $g \raro \chi(g)$ is a character on $SU_2(\IC)$. The group $SU_2(\IC)$ being simply connected, the character is a trivial map. So 
\be 
\psi \beta_{u(g)}=\psi
\ee
for all $g \in SU_2(\IC)$. Since $\tilde{\psi}$ is also an extremal element in $K_{\tilde{\omega}}$, along the same argument we get 
\be 
\tilde{\psi} \beta_{u(g)}=\tilde{\psi}
\ee
for all $g \in SU_2(\IC)$. One can verify (63) directly as well, since 
$\tilde{\psi} \beta_u=\tilde{\psi \beta_{u}}$ 
for any $u \in U_d(\IC)$. Furthermore, for $\psi \beta_u = \psi$, we verify the following steps:
$$\psi_0 (\beta_{u} \otimes \beta_{u}(\tilde{s}_{I'}\tilde{s}^*_{J'} \otimes s_Is_J^*))$$
$$=\phi(\beta_{u}(\tilde{v}_{I'}\tilde{v}_{J'}^*v_Iv_J^*)$$
$$=\phi(\beta_{u}(v_{\tilde{I'}}v_Iv_J^*v^*_{\tilde{J'}}))$$
$$=\phi(v_{\tilde{I'}}v_Iv_J^*v^*_{\tilde{J'}}))$$
$$=\phi(\tilde{v}_{I'}\tilde{v}_{J'}^*v_Iv_J^*)$$
$$=\psi_0(\tilde{s}_{I'}\tilde{s}_{J'}^*s_Is_{J}^*)$$ 
for $|I'|,|J'|,|I|$ and $J| < \infty$. Thus 
\be 
\psi_0 \beta_{u(g)} \otimes \beta_{u(g)} = \psi_0 
\ee 
for all $g \in SU_2(\IC)$. 

\vsp 
We also compute the following elementary equalities for $\psi= \psi \beta_u$ for $u \in U_d(\IC)$:
$$\psi_0 \tilde{\beta}_{u}$$
$$=\psi_0 \beta_u \tilde{\beta}_{_{I_d}}$$
$$=\psi_0 \tilde{\beta}_{I_d}$$
Thus we have 
$$\psi_0 \tilde{\beta}_{u(g)} = \psi_0 \tilde{\beta}_{_{I_d}}$$
for all $g \in SU_2(\IC)$. 

\vsp 
We assume further now that the the state $\omega$ satisfies 
\be 
\omega \beta_{r_0}(\tilde{Q}) = \omega(Q)
\ee 
for all $ Q \in \IM$. Since the state $\omega$ is $G=SU_2(\IC)$-invariant state and $\zeta r_0 \in u(G)$, we have 
$\omega \beta_{r_0}=\omega$ on $\IM$. In particular, the state $\omega$ is lattice symmetric (twist free) state 
of $\IM$ i.e.
\be 
\omega(\tilde{Q})=\omega(Q)
\ee
for all $Q \in \IM$. 

\vsp 
The dual state $\tilde{\psi}$ on $\clo_d$ defined by 
$$\tilde{\psi}(s_Is_J^*)=\psi(s_{\tilde{I}}s^*_{\tilde{J}})$$
for all $|I|,|J| < \infty$ is also an extremal element in $K_{\omega}$. Thus there exists
a $\zeta_0 \in S^1 / H$ so that 
\be 
\tilde{\psi} = \psi \beta_{\zeta_0}
\ee
Since $\tilde{\tilde{\psi}}=\psi$ and $\tilde{\psi \beta_z} = \tilde{\psi} \beta_z$ for any $z \in S^1$, 
we conclude that $\zeta_0^2 \in H$. 

\vsp 
\begin{pro} 
Let $\omega$ be a translation and $SU(2)$-invariant factor state of $\IM$ with (need not be irreducible) a representation $g \raro u(g)$ satisfying (13). Let $\psi$ be an extremal element in $K_{\omega}$ and $\tilde{\psi}$ be the dual state of $\psi$ of $\clo_d$, defined by
$$\tilde{\psi}(s_Is_J^*)= \psi(s_{\tilde{I}} s^*_{\tilde{J}})$$
for all $|I|,|J| < \infty$ and consider the amalgamated state $\psi_0$ on $\tilde{\clo}_d \otimes \clo_d$. Then the following holds:

\NI (a) $\psi \beta_{u(g)} = \psi$ on $\clo_d$;

\NI (b) $\tilde{\psi} \beta_{u(g)} = \tilde{\psi}$ on $\tilde{\clo}_d$;

\NI (c) $\psi_0 \beta_{u(g)} = \psi_0$ on $\tilde{\clo}_d \otimes\clo_d$ for all $g \in SU_2(\IC)$; 

\NI (d) $\psi_0 \beta_{r_{\zeta}} = \psi_0$ on $\tilde{\clo}_d \otimes\clo_d$, where $r_{\zeta} = \zeta r_0 \in u(SU_2(\IC))$.

\vsp 
Let $\omega$ be also reflection symmetric with the twist $r_0$. Then $\omega$ is also reflection symmetric state of $\IM$ and there exists a $\zeta_0 \in S^1 / H$ such that $\zeta_0^2 \in H$ and 

\NI (e) $\tilde{\psi} = \psi \beta_{\zeta_0}$ on $\clo_d$;

\NI (f) $\psi_0 \tilde{\beta}_{u(g)} = \psi_0 \beta_{\zeta_0}$ on $\tilde{\clo}_d \otimes\clo_d$ for all $g \in SU_2(\IC)$;

\NI (g) $\psi_0 \tilde{\beta}_{r_{\zeta}} = \psi_0 \beta_{\zeta_0}$ on $\tilde{\clo}_d \otimes\clo_d$, where $r_{\zeta} = \zeta r_0 \in u(SU_2(\IC))$.

\end{pro} 

\vsp 
\begin{proof} 
We have already proved (a), (b), (c) and (e).  
(d) is a spacial case of (c) with $u(i\sigma_y)=r_{\zeta}$. 

\vsp 
For (f) we verify the following steps:
$$\psi_0 \tilde{\beta}_{u(g)}$$
$$=\psi_0 \beta_{u(g)} \tilde{\beta}_{_{I_d}}$$  
$$=\psi_0 \tilde{\beta}_{_{I_d}}$$
Thus it is good enough, if we verify (f) only for $g=I_d$ as follows:
$$\psi_0 \tilde{\beta}_{_{I_d}}(\tilde{s}_{I'}\tilde{s}^*_{J'} \otimes s_Is_J^*)$$
$$=\psi_0(\tilde{s}_I\tilde{s}^*_{J} \otimes s_{I'}s^*_{J'})$$
$$=\phi(\tilde{v}_{I}v_{I'}v^*_{J'}\tilde{v}^*_{J})$$
$$=\phi(v_{\tilde{I}}v_{I'}v^*_{J'}v^*_{\tilde{J}})$$
$$=\psi(s_{\tilde{I}}s_{I'}s^*_{J'}s^*_{\tilde{J}})$$
$$=\tilde{\psi}(s_{\tilde{I'}}s_Is_J^*s^*_{\tilde{J'}})$$
$$=\psi \beta_{\zeta_0}(s_{\tilde{I'}}s_Is_J^*s^*_{\tilde{J'}})$$
$$=\psi \beta_{\zeta_0}(v_{\tilde{I'}}v_Iv_J^*v^*_{\tilde{J'}})$$
$$=\psi \beta_{\zeta_0}(\tilde{v}_{I'}v_Iv_J^*\tilde{v}^*_{J'})$$
$$=\psi_0 \beta_{\zeta_0}(\tilde{s}^*_{I'}\tilde{s}^*_{J'} \otimes s_Is^*_J)$$ 
for all $|I'|,|J'|,|I|$ and $|J| < \infty$. 

\vsp 
(g) is a spacial case of (f) with $u(i\sigma_y)=r_{\zeta}$. 

\end{proof}

\section{ Real and lattice reflection-symmetric with the twist $r_0$ invariant state }

\vsp
Now we recall from [Mo3] another useful symmetry on $\omega$. If $Q= Q^{(l)}_0 \otimes Q^{(l+1)}_1 \otimes ....\otimes Q^{(l+m)}_m$, we set $Q^t={Q^t}^{(l)}_0 \otimes {Q^t}^{(l+1)}_1 \otimes ..\otimes {Q^t}^{(l+m)}_m$
, where $Q_0,Q_1,...,Q_m$ are arbitrary elements in $\!M_d$ and $Q_0^t,Q^t_1,..$ stands for transpose
with respect to an orthonormal basis $(e_i)$ for $\IC^d$ (not complex conjugate) of $Q_0,Q_1,..$ 
respectively. We define $Q^t$ by extending linearly for any
$Q \in \clb_{loc}$. For a state $\omega$ on $\clb$, we define
a state $\bar{\omega}$ on $\clb$ by the following prescription
\be
\bar{\omega}(Q) =
\omega(Q^t)
\ee
Thus the state $\bar{\omega}$ is a translation-invariant, ergodic, factor state,  if and only if $\omega$ is a translation-invariant, ergodic, factor state respectively. We say $\omega$ is {\it real } if $\bar{\omega}=\omega$. In this section we study a translation-invariant real state.

\vsp
For a $\lambda$ invariant state $\psi$ on $\clo_d$, we define a $\lambda$ invariant state $\bar{\psi}$ as in [Mo3] by
\be
\bar{\psi}(s_Is^*_J)=\psi(s_Js^*_I)
\ee
for all $|I|,|J| < \infty$ and extend linearly. For details, we refer to section 3 in [Mo3].  

\vsp 
\begin{pro} 
Let $\omega$ be a real, lattice symmetric translation invariant pure state of $\IM$. Then there exists an extremal element $\psi$ in $K_{\omega}$, so that $\tilde{\psi}=\psi \beta_{\zeta_0}$ and $\bar{\psi}=\psi \beta_{\zeta_0}$. Let $(\clk,\clm,v_k,1 \le k \le d,\phi)$ be the Popescu element of $\psi$ given as in Proposition 2.4. Then there exists a unique unitary operator $\gamma$ on $\clk$ such that $\gamma \zeta_{\psi}=\zeta_{\psi}$ and 
\be 
\gamma ( \sum c_{I'J',I,J} \tilde{v}_{I'}\tilde{v}_{J'}^*v_Iv^*_J) )\gamma^* = \sum c_{I',J',I,J} \clj \tilde{v}_{I}\tilde{v}_{J}v_{I'}v_{J'}^*\clj
\ee
for all $I'|,|J'|,|I|$ and $|J| < \infty$, where $\gamma$ is also self adjoint, commuting with modular elements $\Delta^{1 \over 2}, \clj$. However, $\gamma u_z = u_{\bar{z}}\gamma$ for all $z \in H$.  

\vsp 
Furthermore, the map $\clj_{\gamma}: \clh \otimes_{\clk} \tilde{\clh} \raro \clh \otimes_{\clk} \tilde{\clh}$, defined by 
\be 
\pi(s_Is^*_J\tilde{s}_{I'}\tilde{s}^*_{J'})\zeta_{\psi} 
\raro \pi(s_{I'}s^*_{J'}\tilde{s}_{I}\tilde{s}^*_{J})\zeta_{\psi}
\ee
for all $\;|I|,|J|,|I'|,|J'| < \infty $ and then extending anti-linearly on their linear span, extends the anti-unitary map $\clj \gamma: \clk  \raro  \clk$ to an anti-unitary map on $\tilde{\clh} \otimes_{\clk} \clh$ so that $\clj_{\gamma}^2=I$ and
$$\pi(\clj_{I_d}(x)) = \clj_{\gamma} \pi(x) \clj_{\gamma}$$ 
for all $x \in \tilde{\clo}_d \otimes \clo_d$. 
\end{pro} 

\vsp 
\begin{proof} 
The state $\omega$ being lattice symmetric and real, as in Theorem 3.4 in [Mo3], we can fix an extremal point $\psi \in K_{\omega}$ such that $\tilde{\psi} = \psi \beta_{\zeta_0}$ and $\bar{\psi}=\psi \beta_{\zeta_0}$, where $\zeta_0 \in \{ 1, e^{i\pi \over n} \}$ and $\zeta_0^2 \in H$. Thus (70) and (71) holds for some unitary operator $\gamma$ on $\clk$ by Theorem 3.4 in [Mo3]. That $\clj_{\gamma}^2=I$ is obvious from the symmetry of the definition.

\vsp 
Since $u_z$ commutes with $\clj$ for $z \in H$, we have the following identities:
$$\gamma u_z v_i u_z^* \gamma^*$$
$$=z \gamma v_i \gamma^*$$
$$=z \clj \tilde{v}_i \clj$$
$$=\clj \bar{z} \tilde{v}_i \clj$$
$$=\clj u_{\bar{z}} \tilde{v}_i u_{\bar{z}}^* \clj$$
$$= u_{\bar{z}} \clj \tilde{v}_i \clj u_{\bar{z}}^*$$
$$=u_{\bar{z}} \gamma v_i \gamma^* u_{\bar{z}}^*$$
for all $1 \le i \le d$. This shows $\gamma^* u_{\bar{z}} \gamma u_z \in \clm'$. Since $\gamma$ and $u_z$ commutes with $\clj$,  $\gamma^* u_{\bar{z}} \gamma u_z \in \clj \clm' \clj = \clm$. $\clm$ being a factor, $\gamma^* u_{\bar{z}} \gamma u_z$ is a scaler multiple of the identity operator on $\clk$. 
However $u_z\zeta_{\psi}=\zeta_{\psi}=\gamma \zeta_{\psi}$ and so the scaler is $1$. Thus we conclude that $u_z \gamma = u_{\bar{z}} \gamma$ for all $z \in H$. 
\end{proof} 

\begin{rem} 
Theorem 3.4 as well as Theorem 3.5 in [Mo3] includes a faulty proof for $\gamma u_z = u_z \gamma$ for $z \in H$. However, the statement with faulty proof is not used in proving the rest of the statements of Theorem 3.4 and Theorem 3.5 in [Mo3]. 
Proof for the main result Theorem 1.3 of the paper [Mo3] did not use the faulty statement.   
\end{rem} 

\vsp 
Let $\omega,\psi$ be as in Proposition 4.1 and $\omega$ be also $SU_2(\IC)$ invariant. By Proposition 3.1, we have $\psi_0 \beta_{u(g)}=\psi_0$ for all $g \in SU_2(\IC)$ on $\tilde{\clo}_d \otimes \clo_d$. Since $r_{\zeta}=u(i\sigma_y)$, there exists a unitary operator $\hat{r}_{\zeta}:\tilde{\clh} \otimes_{\clk} \clh \raro \tilde{\clh} \otimes_{\clk} \clh$ such that 
$\hat{r}_{\zeta} \zeta_{\psi}=\zeta_{\psi}$ and 
$$Ad_{\hat{r}_\zeta}(\pi(x))= \pi(\beta_{r_{\zeta}}(x))$$ 
for all $x \in \tilde{\clo}_d \otimes \clo_d$. 

\vsp 
We recall for each 
$z \in H$, 
$$Ad_{U_z}(\pi(x)) = \pi(\beta_z(x))$$ 
for all $x \in \tilde{\clo}_d \otimes \clo_d$.

\vsp 
We consider the anti-automorphism $\hat{\clj}_{\gamma_{r_{\zeta}}}$ on $\clb(\tilde{\clh} \otimes_{\clk} \clh)$, 
defined by 
$$\hat{\clj}_{\gamma_{r_{\zeta}}}(X) =  \clj_{\gamma} \hat{r}_{\zeta} X 
\hat{r}_{\zeta}^* \clj_{\gamma}$$  
and verify the following 
$$\hat{\clj}_{\gamma_{r_\zeta}}(\pi(\tilde{s}_{I'}\tilde{s}_{J'}s_Is_J^*))$$ 
$$=\clj_{\gamma} \pi(\beta_{r_{\zeta}}(\tilde{s}_{I'}\tilde{s}_{J'}s_Is_J^*))) \clj_{\gamma}$$
$$=\beta_{\bar{r_{\zeta}}}(\clj_{\gamma}\pi(\tilde{s}_{I'}\tilde{s}_{J'}s_Is_J^*)\clj_{\gamma}^*)$$
$$=\beta_{\bar{r_{\zeta}}}(\pi(\clj_{_{I_d}}(\tilde{s}_{I'}\tilde{s}_{J'}s_Is_J^*)))$$
$$=\pi(\clj_{r_{\zeta}}(\tilde{s}_{I'}\tilde{s}_{J'}s_Is_J^*))$$

\vsp 
\begin{pro} 
Let $\omega,\psi$ be as in Proposition 4.1 and $\omega$ be also $SU_2(\IC)$ invariant. Then the anti-automorphism $\clj_{r_{\zeta}}$ on $\tilde{\clo}_d \otimes \clo_d$ induces a well defined anti-automorphism map 
$\hat{\clj}_{r_{\zeta}}$ on $\pi(\tilde{\clo}_d \otimes \clo_d)''$ by 
\be
\hat{\clj}_{r_{\zeta}}(\pi(x)) = \pi(\clj_{r_{\zeta}}(x))
\ee
for all $x \in \tilde{\clo}_d \otimes \clo_d$ and 
$$\hat{\clj}_{r_{\zeta}}=\hat{\clj}_{\gamma_{r_\zeta}}$$ i.e.
\be 
\hat{\clj}_{r_{\zeta}}(X) = \clj_{\gamma} \hat{r}_{\zeta} X \hat{r}^*_{\zeta}\clj_{\gamma}
\ee  
for all $X \in \pi(\tilde{\clo}_d \otimes \clo_d)''$.
Furthermore, $\hat{\clj}_{r_{\zeta}}(P)=P$ and the corner anti-automorphism, defined by 
$$\hat{\clj}_{r_{\zeta}}(a) = P\hat{\clj}_{r_{\zeta}}(PaP)P$$
for all $a \in \clb(\clk)$ satisfies the following: 
\be 
\hat{\clj}_{r_{\zeta}}(a) = \clj \gamma \hat{r}_{\zeta} a  
\hat{r}_\zeta^* \gamma^* \clj 
\ee 
for all $a \in \clb(\clk)$. Furthermore, we have the following consequences: 

\vsp 
\NI (a1) $\hat{\clj}^2_{r_{\zeta}}=\beta_{\mu}$;

\NI (a2) $\beta_{\bar{r_{\zeta}}}(\tilde{S}_I\tilde{S}_J^*S_{I'}S_{J'}^*) \clj_{\gamma} \hat{r_{\zeta}} = \clj_{\gamma}\hat{r_{\zeta}} S_IS_J^*\tilde{S}_{I'}\tilde{S}^*_{J'}$ for all $|I'|,|J'|,|I|$ and $J| < \infty$.

\NI (a3) $Ad_{\hat{U}(g)} \hat{\clj}_{r_{\zeta}}=\hat{\clj}_{r_{\zeta}} Ad_{\hat{U}(g)}$for all $g \in SU_2(\IC)$;

\vsp 
\NI (b1) $Ad^2_{\gamma_{r_{\zeta}}}=\beta_{\zeta^2 I_d}$, where 
$\gamma_{r_{\zeta}}= \gamma \hat{r}_\zeta$ commutes with modular elements $\clj$ and $\Delta^{1 \over 2}$;

\NI (b2) $\beta_{\bar{r_{\zeta}}}(\tilde{v}_I\tilde{v}_J^*v_{I'}v_{J'}^*)  \clj \gamma_{r_{\zeta}} = \clj \gamma_{r_{\zeta}} v_Iv_J^*\tilde{v}_{I'}\tilde{v}^*_{J'}$ for all $|I'|,|J'|,|I|$ and $J| < \infty$;

\NI (b3) $\gamma_{r_{\zeta}}$ commutes the representation $\{\hat{u}(g):g \in SU_2(\IC) \}$;

\vsp 
There exists a unique unitary operator $\Gamma_{\zeta_{r}}$ and an anti-unitary operator extending $\clj$ on $\tilde{\clh} \otimes_{\clk} \clh$ extending unitary $\gamma_{r_{\zeta}}:\clk \raro \clk$ and anti-unitary operator $\clj:\clk \raro \clk$ respectively such that  

\vsp 
\NI (c1) $Ad_{\Gamma_{r_{\zeta}}}^2=\beta_{\zeta^2}$; $Ad_{\Gamma_{r_{\zeta}}}$ acts on $\pi(\clo_d)''$ and $\pi(\mbox{UHF}_d)''$ ( $\pi(\tilde{\clo}_d)''$ and 
$(\pi(\tilde{\mbox{UHF}}_d)''$ ) respectively; 

\NI (c2) $\beta_{\bar{r_{\zeta}}}(\tilde{S}_I\tilde{S}_J^*S_{I'}S_{J'}^*)  \clj \Gamma_{r_{\zeta}} = \clj \Gamma_{r_{\zeta}} S_IS_J^*\tilde{S}_{I'}\tilde{S}^*_{J'}$ 
for all $|I'|,|J'|,|I|$ and $J| < \infty$;

\NI (c3) $\Gamma_{r_{\zeta}}$ and $\clj$ commutes the representation $\{\hat{U}(g):g \in SU_2(\IC) \}$;

\NI (c4) Let $d$ be an even integer then $Ad_{\hat{u}(r_{\zeta})}(x)=x$ if $Ad_{\gamma_{r_{\zeta}}}(x) = x$ for $x \in \clb(\clk)$. Similarly $Ad_{\hat{U}(r_{\zeta})}(X)=X$ if $Ad_{\Gamma_{r_{\zeta}}}(X) = X$ for $X \in  \clb(\clh), \clb(\tilde{\clh})$ or $\clb(\tilde{\clh} \otimes_{\clk} \clh)$.  
\end{pro}

\vsp 
\begin{proof} 
Since $\clj^2_{r_{\zeta}} = \beta_{\bar{r_{\zeta}}r_{\zeta}}= \beta_{\zeta^2 I_{_{d}}}$ on $\tilde{\clo}_d \otimes \clo_d$ by (51), we have 
$$\hat{\clj}_{r_{\zeta}}^2(\pi(x))$$
$$\pi(\clj^2_{r_{\zeta}}(x))$$
$$=\beta_{\zeta^2 I_{_{d}}}(\pi(x))$$ 
for all $x \in \tilde{\clo}_d \otimes \clo_d$. Thus (a1) is true.   

\vsp 
Since $\clj \hat{u}_g = \hat{u}_g \clj$ for all $g  \in SU_2(\IC)$, $\clj$ commutes with $\hat{u}(i\sigma_y)=\hat{\bar{r_{\zeta}}}$. We check the following equalities: 
$$\clj \gamma \hat{\bar{r_\zeta}} \clj \gamma \hat{\bar{r_\zeta}}$$  
$$=\gamma \clj \hat{\bar{r_\zeta}} \clj \gamma \hat{\bar{r_\zeta}}$$
$$=\gamma \clj \clj \hat{\bar{r_{\zeta}}} \gamma \hat{\bar{r_{\zeta}}}$$
$$=\gamma \hat{\bar{r_\zeta}} \gamma \hat{\bar{r_\zeta}}$$
Thus 
$$\hat{\clj}^2_{r_{\zeta}}=Ad^2_{\gamma_\zeta}$$
Since $\hat{\clj}^2_{r_\zeta} = \beta_{\zeta^2 I_{_d}}$, (b1) follows.

\vsp 
For (a2), we recall for any $u \in U_d(\IC)$
$$\clj_{u}(s_k)$$
$$=\beta_{\bar{u}}\clj_{_{I_d}}(s_k) $$
$$=\beta_{\bar{u}}(\tilde{s}_k)$$
Thus 
$$\hat{\clj}_{r_{\zeta}}(\pi(s_k)) = \beta_{\bar{r_{\zeta}}}(\pi(\tilde{s}_k))$$
Now we verify the following simple identities:
$$\beta_{\bar{r_{\zeta}}}(\tilde{v}_k) \clj_{\gamma} \hat{r_{\zeta}}$$
$$=P \beta_{\bar{r_{\zeta}}}(\pi(\tilde{s}_k) P \clj_{\gamma} \hat{r}_{\zeta}P$$
$$=P\pi(\beta_{\bar{r_{\zeta}}}(\tilde{s}_k)\clj_{\gamma} \hat{r}_{\zeta} P$$
$$=P \clj_{\gamma} \hat{r}_{\zeta} \pi(s_k)P$$
$$=P\clj_{\gamma} \hat{r}_{\zeta}P\pi(s_k)P$$
$$=\clj_{\gamma} \hat{r_{\zeta}}v_k$$
The statement (b2) is a simple consequence of (a2). That $\gamma_{r_{\zeta}}$ commutes with modular element is obvious since both $\gamma$ and $\hat{r_{\zeta}}$ commutes with modular elements. 

\vsp 
The statement (a3) is a simple consequence of the inter-twinning relation 
(60) once used in (72). Since $\hat{u}(g)$ commutes with $\clj$ by Proposition 2.7 (c), (b3) is a simple consequence of (a3). 

\vsp 
Now we aim to prove (c1)-(c4). To that end, we consider the minimal Popescu dilation for the elements $(\clj \tilde{v}_i \clj: 1 \le i \le d)$ in $\clm$ to find Cuntz elements say $(T_i:1 \le i \le d )$ acting on a Hilbert space $\clh_T$ such that 
\be 
T_i^*P=PT_i^*P=\clj \tilde{v}^*_i \clj
\ee 
for $1 \le i \le d$ satisfying usual cyclic property as described in Proposition 2.1. Now we also consider its dual Popescu elements i.e. $(\clj v_i \clj)$ in $\tilde{\clm}$ and its minimal dilation to find dual Cuntz elements $(\tilde{T}_i)$ acting on $\tilde{\clh}_T$. We repeat the construction for the amalgamated Hilbert space $\tilde{\clh}_T \otimes \clh_T$ and representation $\pi_T:\tilde{\clo}_d \otimes \clo_d \raro \clb(\tilde{\clh}_T \otimes_{\clk} \clh_T)$ as in section 2. 

\vsp 
Since the unitary operator $\gamma_{\zeta_{r}}$ on $\clk$ inter-twins Popescu elements $(\beta_{r_{\zeta}}(\clj \tilde{v}_i \clj): 1 \le i \le d)$ and $(v_i):1 \le i \le d)$ by (b2), Theorem 5.1 in [BJKW] i.e. Proposition 2.1 (e) ensures a unique unitary operator $\Gamma_{r_{\zeta}}:\clh \raro \clh_T$ which inter-twins Cuntz elements $\{\beta_{r_{\zeta}}(T_i):1 \le i \le d \}''$ and $\{\pi(s_i): 1 \le i \le d \}''$ extending $\gamma_{\zeta_r}:\clk \raro \clk$ i.e. 
there exists a unitary operator $\Gamma_{r_{\zeta}}$ on $\tilde{\clh} \otimes_{\clk} \clh$ 
extending $\gamma_{r_{\zeta}}:\clk \raro \clk$ satisfying 
\be 
\Gamma_{r_{\zeta}}P=P\Gamma_{r_{\zeta}}P=\gamma_{r_{\zeta}}
\ee 
and 
\be 
\beta_{r_{\zeta}}(\pi_T(x)) \Gamma_{r_{\zeta}} =  \Gamma_{r_{\zeta}} \pi(x)
\ee  
for $x \in \tilde{\clo}_d \otimes \clo_d$. Thus without a loss of generality, we assume that $\clh_T=\clh$ and the von-Neumann algebras $\{T_i:1 \le i \le d \}''$ and $\{\pi(s_i): 1 \le i \le d \}''$ are equal. Along the same line of argument, we identify $\tilde{\clh}_T$ with $\tilde{\clh}$ and so $\tilde{\clh}_T \otimes_{\clk} \clh_T$ with $\tilde{\clh} \otimes_{\clk} \clh$. 

\vsp 
Since $\gamma_{r_{\zeta}}^2=\beta_{\zeta^2I_d}$, the unitary operator $\gamma^*_{r_{\zeta}}$ also inter-twins Popescu elements $\beta_{\bar{r}_{\zeta}}(\clj \tilde{v}_i \clj)$ and $(\beta_{\zeta^2}(v_i))$, by uniqueness of commutant theorem $\Gamma_{\zeta_{r}}^*= U_{\zeta^2} \Gamma_{\zeta_r}$ i.e.  
\be 
\Gamma_{r_{\zeta}}^2 = \beta_{\zeta^2}
\ee 

\vsp 
That $Ad_{\Gamma_{r_{\zeta}}}$ acts on $\pi(\clo_d)''$ follows by construction. That it also acts on $\pi(\mbox{UHF}_d)''$ 
follows by Proposition 2.2 (b) and $Ad_{\Gamma_{r_{\zeta}}} \beta_z = \beta_{\bar{z}} Ad_{\Gamma_{r_{\zeta}}}$ for all 
$z \in H$. So by our construction $Ad_{\Gamma_{r_{\zeta}}}$ acts on $\pi(\clo_d)''$ and $\pi(\tilde{\clo}_d)''$ respectively with 
$$Ad_{\Gamma_{r_{\zeta}}}^2= Ad_{\Gamma_{r_{\zeta}}^2}= \beta_{\zeta^2}$$

\vsp 
Furthermore, by the commutant lifting theorem Proposition 2.1 (e), we also find an anti-unitary operator $\clj: \tilde{\clh} \otimes_{\clk} \clh  \raro \tilde{\clh} \otimes_{\clk} \clh$, extending the anti-unitary map $\clj: \clk \raro \clk$, such that 
\be  
\clj \pi_T(x) = \pi(\clj_{I_d}(x)) \clj
\ee   
for all $x \in \tilde{\clo}_d \otimes \clo_d$. 

\vsp 
Now we use (77) with (79) to get 
\be 
\beta_{\bar{r}_{\zeta}}(\pi(\tilde{s}_{I'}\tilde{s}_{j'}s_Is_J^*))\clj \Gamma_{r_{\zeta}} = 
\clj \Gamma_{r_{\zeta}} \pi(\tilde{s}_{I}\tilde{s}_{J}s_{I'}s^*_{J'})  
\ee
for all $|I'|,|J'|,|I|$ and $|I| < \infty$. Now it is evident by (73) and (80) that 
\be 
\clj \Gamma_{r_{\zeta}} =  \clj_{\gamma} \hat{r}_{\zeta}
\ee 
Since $(\clj_{\gamma} \hat{r}_{\zeta})^2= U_{\zeta^2}$, using (81) we get 
$$\clj \Gamma_{r_{\zeta}} $$
$$=\Gamma_{r_{\zeta}}^* \clj U_{\zeta^2}$$
$$=\Gamma_{r_{\zeta}} \clj$$
where we have used $\clj^2=I$ and $\Gamma_{r_{\zeta}}^2=U_{\zeta^2}=\hat{r}_{\zeta}^2$. 

\vsp 
Proposition 2.7 is valid with a representation $g \raro \hat{U}_T(g)$ on $\tilde{\clh} \otimes_{\clk} \clh$ so that 
\be 
\hat{U}_T(g) T_i \hat{U}_T(g)^*= \bar{\chi(g)} \beta_{\bar{u}(g)}(T_i)
\ee
for all $1 \le i \le d$. Thus 
$$\clj \hat{u}_T(g) \tilde{v}_i \hat{u}_T(g)^* \clj$$
$$=\hat{u}_T(g) \clj \tilde{v}_i \clj \hat{u}_T^*(g)$$
$$= \bar{\chi(g)} \beta_{\bar{u}(g)}(\clj \tilde{v}_i\clj)$$
$$=\clj \chi(g) \beta_{u(g)}(\tilde{v}_i) \clj$$
for all $1 \le i \le d$ and $g \in G$. Thus the representation $g \raro \hat{u}_T(g)$ as well satisfies 
the covariance relation (44) that the representation $g \raro u(g)$ satisfies. In other words, $\hat{u}(g)^*\hat{u}_T(g) \in \tilde{\clm}'=\clm$. Since $\clj$ commutes with both $\hat{u}(g)$ and $\hat{u}_T(g)$, $\hat{u}(g)^*\hat{u}_T(g)$ is also an element in $\clj \clm \clj=\clm'$ i.e. $\hat{u}(g)^*\hat{u}_T(g)$ is an element in the centre of $\clm$. Since $\clm$ is a factor and $\hat{u}(g)\zeta_{\psi}=\zeta_{\psi}=\hat{u})g)\zeta_{\psi}$, we get 
\be 
\hat{u}(g)=\hat{u}_T(g)
\ee 
for all $g \in SU_2(\IC)$. Furthermore, by the covariant relations (43) and (82) used in (79), we also have 
\be 
\clj \hat{U}_T(g) = \hat{U}(g) \clj
\ee 
for all $g \in SU_2(\IC)$. 

\vsp 
We fix any $g \in SU_2(\IC)$. We consider unitary operators $U(g)$ and $U_T(g)$ restricting its action 
on $\clh$. We claim that $\hat{U}(g)=\hat{U}_T(g)$ on $\clh$ for all $g \in SU_2(\IC)$. For a proof we recall the commutant lifting theorem ( Theorem 5.1 in [BJKW] ) in a little more details. The unitary 
operator 
\ben
U_2(g) = \left (\begin{array}{llll} 0&,&\; \hat{U}(g) \\ \hat{U}^*(g) &,&\;\;0 \\ 
\end{array} \right ),
\een
is the unique element in the commutant of the Cuntz algebra generated by 
\ben
\left (\begin{array}{llll} S_i&,&\; 0 \\ 0&,&\;\;\beta_g(S_i) \\ 
\end{array} \right ),
\een
such that 
\be
P_2U_2(g)P_2 = \left (\begin{array}{llll} 0&,&\; \hat{u}(g) \\ \hat{u}(g)^*&,&\;\;0 \\ 
\end{array} \right ),
\ee
where 
\ben
P_2 = \left (\begin{array}{llll} P&,&\; 0 \\ 0&,&\;\;P \\ 
\end{array} \right ),
\een
Similar statement also holds for $\hat{U}_T(g)$ with Cuntz elements $(T_i)$ replacing the role of $(S_i)$
above. Since $\pi_T(\clo_d)''=\pi(\clo_d)''$ and the element 
\ben
\left (\begin{array}{llll} 0&,&\; \hat{U}_T(g) \\ \hat{U}_T^*(g) &,&\;\;0 \\ 
\end{array} \right ),
\een
also satisfies (85) as $\hat{u}_T(g)=\hat{u}(g)$, the uniqueness part of the above 
statement says that $\hat{U}(g)=\hat{U}_T(g)$ on $\clh$ by Proposition 2.1 (d). By the same argument, 
we also have $\hat{U}(g)=\hat{U}_T(g)$ on $\tilde{\clh}$. This shows $\hat{U}(g)=\hat{U}_T(g)$ and $\hat{U}(g)$ 
commutes with $\clj$. Since $\hat{U}(g)$ commutes with $\clj_{\gamma}\hat{r}_{\zeta}$, we conclude $\hat{U}(g)$ 
commutes with $\Gamma_{r_{\zeta}}$ by (81)
and the commuting property of $\hat{U}(g)$ with $\clj$. This completes the proof for (c3). 

\vsp 
We fix any even integer $d$. We recall that $r_{\zeta}=\hat{u}(i\sigma_y)$. The fixed point von-Neumann sub-algebra of 
the action $Ad_{r_{\zeta}}$ on a von-Neumann algebra is equal to the fixed point von-Neumann sub-algebra of the group 
action $\{ Ad_{\hat{u}(e^{it\sigma_y})}:0 \le t < 2\pi \}$ on $\clb(\clk)$ since $e^{it\sigma_y}$ takes value 
$i\sigma_y$ at $t={\pi \over 2}$. We used here the simple fact that the fixed point sub-algebra remains same if we 
take action of a single element from the group other than the identity action by von-Neumann double commutant theorem. 
We also note that $r^2_{\zeta}=\hat{u}(i^2\sigma^2_y)= \hat{u}(-I_2)$. Important point here for even values of $d$, 
$\hat{u}(-I_2)$ is not the identity operator. So the fixed point algebra of the action $Ad_{r_{\zeta}}$ is equal to 
the fixed point von-Neumann sub-algebra of the action $Ad_{\hat{u}(-I_2)}$ i.e. the action at $t=\pi$. 

\vsp 
In particular, we claim that any element in the fixed point von-Neumann sub-algebras of the actions $Ad_{\gamma_{r_{\zeta}}}$ 
on $\clb(\clk)$ is an invariant element of $Ad_{\hat{u}_{r_{\zeta}}}$ on $\clb(\clk)$. For a proof, we take any element $a \in \cla$ fixed by $Ad_{\gamma_{r_{\zeta}}}$. 
Then we get 
$$a$$
$$=Ad^2_{\gamma_{r_{\zeta}}}(a)$$
$$=Ad_{\gamma^2_{\zeta}}(a)$$
$$=\beta_{\zeta^2I}(a)$$
$$=Ad_{\hat{u}^2(i\sigma_y)}(a)$$
$$=Ad_{\hat{u}(-I_2)}(a)$$
Now we use the preceding remark to conclude that $Ad_{\hat{u}(r_{\zeta})}(a)=a$. 

\vsp 
We use the same argument to prove $Ad_{\hat{U}(r_{\zeta})}(X)=X$ if $Ad_{\Gamma_{r_{\zeta}}}(X)=X$ 
for some $X \in \clb(\tilde{\clh} \otimes_{\clk} \clh)$. 
\end{proof} 

\vsp 
In the following we aim to give a more insight in our studies on symmetries on state $\omega$. 

\vsp 
\begin{pro} 
Let $\omega$ and $\psi$ be as in Proposition 4.1 with $\tilde{\psi}=\bar{\psi}=\psi \beta_{\zeta_0}$. Let $\omega$ be also $\beta_{r_0}$ invariant then the following holds:

\NI (a) $\psi \beta_{r_0} = \psi \beta_{\eta_0}$ and $\tilde{\psi} \beta_{r_0} = \tilde{\psi} \beta_{\eta_0}$ for some $\eta_0 \in S^1$;

\NI (b) Let $\eta \in S^1$ such that $\eta^2=\eta_0$ and $\psi_{\eta}=\psi \beta_{\eta}$. Then 
$\tilde{\psi_\eta} \beta_{r_0} = \psi_{\eta} \beta_{\eta_0\zeta_0}$ and $\bar{\psi_{\eta}} = \psi_{\eta} \beta_{\eta_0 \zeta_0}$;

\NI (c) Let $(l_k=\eta v_k)$ and then $(\tilde{l}_k=\eta \tilde{v}_k)$ be the Popescu elements of the states $\psi_{\eta}$ and $\tilde{\psi_{\eta}}$ of $\clo_d$ and $\tilde{\clo}_d$ respectively in their support projections. Then there exists a unitary operator $\gamma_{r_0}$ on $\clk$ such that 
\be 
\gamma_{r_0}\psi_{\zeta}=\psi_{\zeta},\;\;\gamma_{r_0}\beta_{\bar{r_0}} (\tilde{l}_{I'}\tilde{l}^*_{J'}l_Il_J^*)\gamma_{r_0}^* = \clj \tilde{l}_I\tilde{l}_J^*l_{I'}l_{J'}^*\clj, 
\ee  
and a unitary operator $\hat{r}_0$ on $\clk$ satisfying 
\be 
\hat{r}_0\tilde{l}_{I'}\tilde{l}_{J'}^*l_Il_J^*\hat{r}_0^*=\beta_{r_0}(\tilde{l}_{I'}\tilde{l}_{J'}^*l_Il_J^*)
\ee
for all $|I'|,|J'|,|I|$ and $|J| < \infty$. Moreover, $\gamma_{r_0}u_z =u_{\bar{z}}\gamma_{r_0}$ for all $z \in H$;

\NI (d) If $\omega$ is also $SU_2(\IC)$ invariant as in Proposition 4.3 then $\zeta \in H$ and (a)-(c) are valid with 
$\eta=\eta_0=1$. 
\end{pro} 

\begin{proof} 
We have already fixed an extremal point $\psi$ satisfying $\tilde{\psi} = \psi \beta_{\zeta_0}$ and $\bar{\psi}=\psi \beta_{\zeta_0}$ as in Proposition 4.1. The state $\psi \beta_{r_0}$ of $\clo_d$ being an extremal point in $K_{\omega}$, there exists a $\eta_0 \in S^1$ such that $\psi \beta_{r_0}=\psi \beta_{\eta_0}$. Since $r_0^2=I_d$, $\eta_0^2 \in H$. 
Since $\tilde{\psi}=\psi \beta_{\zeta_0}$ and $\beta_{\zeta_0}$ commutes with $\beta_{r_0}$, we check 
that 
$$\tilde{\psi} \beta_{r_0}$$
$$=\psi \beta_{\zeta_0} \beta_{r_0}$$
$$=\psi \beta_{r_0} \beta_{\zeta_0}$$
$$=\psi \beta_{\eta_0} \beta_{\zeta_0}$$
$$=\psi \beta_{\zeta_0} \beta_{\eta_0}$$
$$=\tilde{\psi} \beta_{\eta_0}$$
and 
$$\tilde{\psi} \beta_{r_0} = \psi \beta_{\zeta_0 \eta_0}$$
We set $$\psi_{\eta} = \psi \beta_{\eta},$$ 
where $\eta^2=\eta_0$.

\vsp 
Since $\tilde{\psi \beta_z} = \tilde{\psi} \beta_z$ for all $z \in S^1$, we have and verify the following equalities: 
$$\tilde{\psi_{\eta}} \beta_{r_0} $$
$$= \tilde{\psi} \beta_{\eta} \beta_{r_0}$$   
$$= \tilde{\psi} \beta_{r_0} \beta_{\eta}$$
$$= \psi \beta_{\eta \zeta_0 \eta_0}$$
$$= \psi_{\eta} \beta_{\zeta_0\eta_0} $$

\vsp 
We also verify the following equalities: 
$$\bar{\psi_{\eta}} $$
$$= \bar{\psi} \beta_{\bar{\eta}} $$
$$= \psi \beta_{\zeta_0} \beta_{\bar{\eta}}$$
$$= \psi \beta_{\eta} \beta_{\zeta_0} \beta_{\bar{\eta}_0}$$
$$= \psi_{\eta} \beta_{\zeta_0 \eta_0}$$
since $\eta_0^2 \in H$. 

\vsp 
So by Theorem 3.4 in [Mo3] we find unitary operator $\gamma_{r_0}$ on $\clk$ satisfying (86). In particular, 
(86) shows that $\gamma_{r_0}$ is a unitary operator inter-twinning the Popescu elements
$(\clj \beta_{\bar{r_0}}(\tilde{l}_k) \clj)$ and $(l_k)$. Let $\pi_T$ be the minimal Popescu dilation
associated with $(\clj \tilde{v}_i \clj)$ and its extended amalgamated representation of $\tilde{\clo}_d \otimes \clo_d$ 
as in Proposition 4.3. Then as in Proposition 4.3, we use commutant lifting theorem to find a unique unitary operator 
$\Gamma_{r_0}$ on $\tilde{\clh} \otimes_{\clk} \clh$ such that 
$$P \Gamma_{r_0} P = \gamma_{r_0},\; \beta_{r_0}(\pi_T(\beta_{\eta}(x))) \Gamma_{r_0} = \Gamma_{r_0} \pi(\beta_{\eta} (x))$$
Since $\Gamma_{r_0} P \Gamma_{r_0} = P$ and the automorphism that maps 
$$\pi(\beta_{\eta}(s_i)) \raro \Gamma_{r_0}^* \pi_T(\beta_{\eta}(s_i)) \Gamma_{r_0} = \pi(\beta_{r_0}(\beta_{\eta}(s_i)))$$
gives a well defined automorphism on $\clb(\clk)$ that takes 
$$\tilde{l}_{I'}\tilde{l}_{J'}^*l_Il_J^* \raro \beta_{r_0}(\tilde{l}_{I'}\tilde{l}_{J'}^*l_Il_J^*)$$
for all $I'|,|J'|,|I|$ and $|J| < \infty$. Since $\clm \vee \tilde{\clm}=\clb(\clk)$ by Proposition 2.4 (c), we conclude that 
there exists a unitary operator $\hat{r_0}$ on $\clk$ satisfying (87) by a standard result. Note that we are not claiming that $\hat{r}_0\zeta_{\psi}=\zeta_{\psi}$!

\vsp 
By the $SU_2(\IC)$-invariance property of $\omega$, we have $\psi \beta_{u(g)}=\psi$ for all $g \in SU_2(\IC)$ 
by Proposition 3.1 and so in particular $\psi \beta_{r_{\zeta}}=\psi$. Since $r_{\zeta}=\zeta r_0$ and so 
$\beta_{\zeta}= \beta_{r_{\zeta}}\beta_{r_0}$ on $\clo_d$ has a normal extension to $\pi_{\psi}(\clo_d)''$, given by $Ad_{r_{\zeta}}Ad_{r_0}$. Thus by Proposition 2.2 (a) $\zeta \in H$ and $\hat{r}_0\zeta_{\psi}=\zeta_{\psi}$.  
\end{proof}

\vsp 
\begin{thm}
Let $\omega$ be a translation invariant pure state $\omega$ of $\IM=\otimes_{k \in \IZ } \!M^{(k)}_{d}(\IC)$ with the 
following properties: 

\NI (a) $\omega$ is real and lattice reflection symmetric with the twist $r_0 \in U_d(\IC)$, 
where $r_0^2=I_{d}$;

\NI (b) $\omega = \omega \beta_{u(g)}$ for all $g \in SU_2(C)$, where $g \raro u(g)$ is the irreducible representation satisfying the inter-twining relation (55) with the twist $r_0$, where $r_0 \bar{r}_0 = \zeta^2 I_{d}$ and $\zeta^2=1$ for odd integer and $-1$ for even integer. 

\vsp 
If $\omega$ is also reflection positive with twist $r_0$ then $d$ is an odd integer i.e. there exists no translation invariant pure state of $\IM$ satisfying (a), (b) and reflection positive with twist $r_0$ for even values of $d$. 
\end{thm} 

\vsp 
\begin{proof} 

\vsp 
We consider the group action $Ad_{\Gamma_{r_{\zeta}}}$ on the von-Neumann algebra $\pi(\clo_d)''$. Since
$\beta_z Ad_{\Gamma_{r_{\zeta}}}=Ad_{\Gamma_{r_{\zeta}}} \beta_{\bar{z}}$ for all $z \in H$ on $\pi(\clo_d)''$ and $\beta_z$ invariant elements in $\pi(\clo_d)''$ and $\pi(\tilde{\clo}_d)''$ are $\pi(\mbox{UHF}_d)''$ and $\pi(\tilde{\mbox{UHF}}_d)''$ respectively by Proposition 2.4,$Ad_{\Gamma_{r_{\zeta}}}$ acts on $\pi(\mbox{UHF}_d)''$ and $\pi(\tilde{\mbox{UHF}}_d)''$ respectively. In particular, $Ad_{\Gamma_{r_{\zeta}}}$ acts on $\pi(\tilde{\mbox{UHF}}_d \otimes \mbox{UHF}_d)''$. By a theorem of R. T. Powers [Pow], there exits a unique automorphism $Ad_{\Gamma_{r_{\zeta}}}$ on $\tilde{\mbox{UHF}}_d \otimes \mbox{UHF}_d$ such that
$$Ad_{\Gamma_{r_{\zeta}}}(\pi(x))= \pi(Ad_{\Gamma_{r_{\zeta}}}(x))$$
It is a routine work to verify by the uniqueness part of the above statement that 
$\Gamma_{r_{\zeta}}$ commutes with $Ad_{\hat{U}(r_{\zeta})}$ on $\tilde{UHF}_d \otimes \mbox{UHF}_d$. 

\vsp 
So far we did not use our assumption on $d$. We will prove that $d$ is an odd integer by bringing a contradiction for even integer values of $d$. Let $d$ be an even integer. If $Ad_{\Gamma_{r_{\zeta}}}(x)=x$ for $x \in \tilde{\mbox{UHF}}_d \otimes \mbox{UHF}_d$ 
then $Ad_{\hat{U}(r_{\zeta})}(x)=x$ by Proposition 4.3 (c4). The fixed point algebra of $Ad_{\Gamma_{r_{\zeta}}}$ 
being a sub-algebra of a simple $C^*$ algebra $\tilde{\mbox{UHF}}_d \otimes \mbox{UHF}_d$, it is also simple. The automorphism  $Ad_{u(\hat{r}_{\zeta})}$ commutes with $Ad_{\Gamma_{r_{\zeta}}}$ on $\pi(\tilde{\mbox{UHF}}_d \otimes \mbox{UHF}_d)''$, we get 
by a Robert's version [BE] of `Tanaka duality theorem', $Ad_{\hat{U}(r_{\zeta})}$ is one of the element in the group generated by $Ad_{\Gamma_{r_{\zeta}}}$. For details, we refer to Corollary 4.6 in [BE] valid for a more general situation. Since $Ad_{u(r_{\zeta})}$ is not equal to the identity automorphism on $\!M_d(\IC)$, the automorphism $Ad_{\hat{U}(r_{\zeta})}$ 
is either $Ad_{\Gamma_{r_{\zeta}}}$ or $Ad_{\Gamma^*_{\zeta}}$ on $\pi(\mbox{UHF}_d)''$. In particular, $Ad_{\hat{u}(r_{\zeta})}$ is either $Ad_{\gamma_{r_{\zeta}}}$ or $Ad_{\gamma^*_{r_{\zeta}}}$ or $\clm_0$, where we recall $\clm_0=P\pi(\mbox{UHF}_d)''P$ in
Proposition 2.4.

\vsp 
But for a reflection positive with the twist $r_0$ state $\omega$ satisfying (a) and (b), we have $Ad_{\gamma_{r_{\zeta}}}(a)=Ad_{\gamma_0}(a)=a$ for all $a \in \clm_0$ by Theorem 3.5 (d) in [Mo3]. This brings a contradiction to our assumption 
that $d$ is an even integer.  
\end{proof} 

\vsp 
In the following, we remove the additional assumption `reflection positivity with the twist $r_0$' on the state $\omega$ in 
Theorem 4.5. This came as a surprise!

\vsp 
\begin{thm} 
There exists no translation invariant pure state $\omega$ on $\IM = \otimes_{ k \in \IZ} \IM^{(k)}_d(\IC)$
satisfying (a) and (b) in Theorem 4.5 if $d$ is an even integer.
\end{thm} 

\begin{proof} 

\vsp 
To that end, we consider automorphism $Ad_{\Gamma_{r_{\zeta}}}$ that acts on $\pi(\clo_d)''$ and consider
the $C^*$ sub-algebra $\cla= \pi(\clo_d) \bigcap Ad_{\Gamma_{r_{\zeta}}}(\pi(\clo_d))$. It is clear that 
$\cla \subseteq \pi(\clo_d)''$ and $\cla' = \pi(\clo_d)' \vee Ad_{\Gamma_{r_{\zeta}}}(\pi(\clo_d)') = 
\pi(\clo_d)'$ since $Ad_{\Gamma_{r_{\zeta}}}(\pi(\clo_d)') = Ad_{\Gamma_{r_{\zeta}}}(\pi(\clo_d))'=\pi(\clo_d)'$ 
i.e. $\cla'' = \pi(\clo_d)''$. 

\vsp 
Since $Ad_{\Gamma_{r_{\zeta}}}^2= \beta_{\zeta^2}$, the automorphism $Ad_{\Gamma_{r_{\zeta}}}$ acts on $\cla$. 
Since $Ad_{\hat{U}(r_{\zeta})}$ commutes with $Ad_{\Gamma_{r_{\zeta}}}$ on $\pi(\clo_d)''$ and $Ad_{\hat{U}(r_{\zeta})}$ 
acts on $\pi(\clo_d)$, we verify that $Ad_{\hat{U}(r_{\zeta})}$ as well acts on $\cla$ and 
the action commutes with $Ad_{\Gamma_{r_{\zeta}}}$ on $\cla$. The $C^*$-algebra $\cla$ is $C^*$-sub-algebra of a simple 
$C^*$-algebra $\clo_d$, $\cla$ is itself simple. In particular, fixed point sub-algebra $\cla_{Ad_{\Gamma_{r_{\zeta}}}}$ is also simple. So we may once again appeal to Corollary 4.6 in [BE] to conclude that $Ad_{\hat{U}(r_{\zeta}}=Ad_{\Gamma_{r_{\zeta}}}$ or $Ad_{\Gamma^*{r_{\zeta}}}$ 
on $\cla$. Since $\cla''=\pi(\clo_d)''$, by uniqueness of the normal extension, we get 
$Ad_{\hat{U}(r_{\zeta})}=Ad_{\Gamma_{r_{\zeta}}}$ or $Ad_{\Gamma^*{r_{\zeta}}}$ on $\pi(\clo_d)''$. 
Since $Ad_{\Gamma_{r_{\zeta}}}(U_z)=U_{\bar{z}}$ and $Ad_{\hat{U}(r_{\zeta})}(U_z)=U_z$ for all $z \in H$, 
we get $z^2=1$ for all $z \in H$ i.e. $H \subseteq \{ z \in S^1: z^2=1 \}$. Furthermore, 
we have 
\be
S_i = \clj \beta_{\epsilon}(\tilde{S}_i) \clj 
\ee
with $\epsilon=1$ or $-1$. Now we use the commuting property of $\hat{U}(g)$ with $\clj$ in (88) to conclude 
$u^i_j(g) = \overline{u^i_j(g)}$ for all $g \in SU_2(\IC)$ by the covariance relation (43). 
Thus by (55), $r_{\zeta}$ commutes with the irreducible representation $g \raro u(g)$ on $\IC^d$. This brings 
a contradiction since $r_{\zeta}$ is not a scaler multiple of the identity operator $I_d$. 

\vsp 
A simpler alternative proof goes as follows. By Proposition 4.4 (d), $\zeta \in H$ and so $\zeta^2=1$ by the first part 
of the argument used for even values of $d$. This brings a contradiction as $\zeta^2=-1$ for even values of $d$. 
This completes a proof that $d$ can not be an even integer. 
\end{proof}

\vsp 
We have the following generalisation of Theorem 4.6.

\vsp 
\begin{thm} 
Let $\omega$ be a translation invariant pure state of 
$$\IM(n) = \otimes_{\ul{j} \in \IZ \otimes \IZ_n} \!M_d^{(\ul{j})}(\IC) = \otimes_{j \in \IZ} \otimes_{1 \le l \le n} \IM^{(j,l)}_d(\IC)$$ 
satisfying (a) and (b) in Theorem 4.5 with $\ul{r_0} = \otimes r^{(\ul{l})}_0$ and $\ul{u(g)} = \otimes_{1 \le l \le n} u(g)^{(\ul{l})}$. If $d$ is an even integer then $n$ is an even integer.  
\end{thm} 

\vsp 
\begin{proof}
Since $r_0\bar{r}_0=-1$ for even values of $d$, we get $\beta_{\ul{r_{0}}}^2=\beta_{(-1)^nI}=\beta_{-I}$ on 
$$\IM(n) \equiv \otimes \IM^{(k)}_{dn}(\IC)$$ 
for odd values of $n$. So by the same argument used in the proof for Theorem 4.6, $n$ can not be an odd integer.    
\end{proof} 

\vsp 
We end this section stating a result that explains how $SU_2(\IC)$ invariant is crucial in the proof of Theorem 4.5. We consider the following standard ( irreducible ) representation of Lie algebra $su_2(\IC)$ in $\IC^2$:
\ben
\sigma_x = \left (\begin{array}{llll} 0&,&\; 1 \\ 1&,&\;\;0 \\ 
\end{array} \right ),
\een
\ben
\sigma_y = \left (\begin{array}{llll} \;0&,&\;\; i \\ -i&,&\;\;0 \\
\end{array} \right ),
\een
\ben
\sigma_z = \left (\begin{array}{llll} 1&,&\;\; 0 \\  0&,& \;-1\\
\end{array} \right ).
\een
\ben
r_0 = \left (\begin{array}{llll} \;0&,&\;\; i \\ -i&,&\;\;0 \\
\end{array} \right ),
\een

\vsp
Let $\omega$ be a translation invariant pure state on $\IM=\otimes_{\IZ}\!M_2(\IC)$. If $\omega$ is $G=U(1) \subseteq SU_2(\IC)$ invariant then by 
a theorem of T. Matsui [Ma3], $\omega$ is either a product state or a non-split state. The unique ground state for $H_{XY}$ model is a non spilt state.. The following corollary says more when $\omega$ is also real and lattice symmetric.    

\vsp 
\begin{cor} 
Let $\omega$ be translation invariant pure state of $\IM= \otimes_{k \in \IZ} \!M^{(k)}_{2}(\IC)$ and $\omega$ be also lattice reflection symmetric, lattice reflection symmetric with twist $r_0$ and real. If $\omega$ is also $S^1 \subset SU_2(\IC)$ invariant and reflection positive with the twist $r_0$ then $\{u_z: z \in H \}$ does not commutes with $\gamma_{r_0}$ i.e. $H$ is not a subset of $\{-1,1\}$. In such a case, spatial correlation function of $\omega$ does not decay exponentially.  
\end{cor} 

\vsp 
\begin{proof} 
Suppose $\gamma_{r_0}$ commutes with $\{u_z:z \in H \}$. The fixed point sub-algebra of group action $\{Ad_{u_z}: z \in H \}$ is $\clm_0$ by Remark 2.5. The action  $Ad_{\gamma_{r_0}}$ also keeps $\clm_0$ fixed by the reflection positivity property as shown in Theorem 3.5 in [Mo3]. Thus we conclude that $Ad_{\gamma_{r_0}}$ is equal to $Ad_{u_{z_0}}$ for some $z_0 \in H$ by a version of `Tanaka duality theorem' [BE], where we have also used Remark 2.6 that $\clm_0$ is a factor for pure state $\omega$. Since $Ad_{\gamma_0}^2=\beta_{-I_2}$, we get $z_0^2=-1$. Thus $i \in H$.  

\vsp 
On the other hand, by Proposition 4.4 (c), $\gamma_{r_0}u_z=u_{\bar{z}}\gamma_{r_0}$ for all $z \in H$. Thus $\gamma_{r_0}$ commutes 
with $\{u_z: z \in H \}$, if and only if $H \subseteq \{1,-1\}$. This brings a contradiction.      

\vsp
For such a state $\omega$, Theorem 1.3 in [Mo3] says that the exponential decaying property of the spatial correlation 
function of $\omega$ is equivalent to split property of $\omega$ i.e. $\pi_{\omega}(\IM_R)''$ is a type-I factor. If so, 
then by Proposition 2.2 (d), $H=\{1\}$. Since $H$ is not equal to $\{1\}$ by the first part, we get a contradiction. This
completes the proof. 
\end{proof}

\vsp 
\begin{rem} 
In the above corollary, since $Ad_{\gamma_{r_0}}^2=\beta_{-1}$, by Proposition 2.2 (a) $-1 \in H$. Thus 
$\{-1, 1 \} \subset H $ with strict inclusion.     
\end{rem}

\vsp 
\begin{cor} 
Let $d$ be an odd integer and $\omega$ be a reflection positive translation invariant state with twist $r_0$ satisfying (a) and (b) in Theorem 4.5 and $\psi$ be as in Proposition 4.1. If $H \subseteq \{-1, 1 \}$ then 
\be 
\clj \beta_{\epsilon}(\tilde{v}_i) \clj = \beta_{r_0}(v_i),
\ee
where $\epsilon$ is either $1$ or $-1$. If $H=\{1\}$ then $\epsilon=1$ in (89).  
\end{cor} 

\vsp 
\begin{rem} 
The unique ground state for AKLT model [AKLT] is an example that satisfies criteria of 
Corollary 4.10 with $H=\{1\}$. 
\end{rem}

\section{Spontaneous symmetry breaking in ground states of Hamiltonian in quantum spin chain }

\vsp 
We are left to discuss few motivating examples for this abstract framework we have developed so far to study symmetry of Hamiltonian $H$ that satisfies (3) and (14). Before we take specific examples, we recall some well known results in the following proposition for our reference.     

\vsp 
\begin{pro} 
Let $H$ be a Hamiltonian in quantum spin chain $\IM= \otimes_{\IZ} \!M_d(\IC)$ 
that satisfies relation (3) with $h_0 \in \IM_{loc}$. Then the following statements are true:

\NI (a) There exists a unique KMS state $\omega_{\beta}$ for $(\alpha_t)$ at each inverse positive temperature $\beta={1 \over kT} > 0$;

\NI (b) If $H$ also satisfies relation (14) with $J > 0$ and $r_0 \in U_d(\IC)$, then the unique KMS state $\omega_{\beta}$ is reflection positive with twist $r_0$. Furthermore, any weak$^*$ limit point of $\omega_{\beta}$ as $\beta \raro \infty$ is also reflection positive with twist 
$r_0$;

\NI (c) If ground state of $H$ that satisfies (3) and (14) with $J > 0$ is unique, then the unique ground state is pure, translation invariant and reflection positive state with twist $r_0$.   
\end{pro}

\begin{proof} 
For (a), we refer to H. Araki work [Ara2] and also [Ki]. For the first statement in (b), we refer to [FILS]. Last part of (b) is trivial as reflection positive property (11) is closed under weak$^*$ limit. (c) follows by a simple application of (b) since unique ground state is a limit point of positive temperature states.   
\end{proof} 

\begin{proof} 
(Corollary 1.3) We will prove the statement by bringing a contradiction. Suppose that ground state is unique. Then unique ground state of $H$ will inherit 
all symmetries of $H$ i.e. translation, real, lattice symmetric properties of $H$ as it is the liming state of unique $\beta-$KMS states as $\beta \raro \infty$ and each $\beta-$KMS state being unique, inherits these symmetries of $H$. That it is also reflection positive with the twist $r_0$ follows along the same line, once we check the limit of refection positive states with the twist $r_0$ remains reflection positive with the twist $r_0$ by (8). This brings a contradiction to Theorem 4.1 since the ground state of $H$ is pure being unique.  
\end{proof} 

\begin{proof} 
(Corollary 1.4) Since the invariant subspace $E_{\omega}=\{f \in \clh_{\omega}: e^{itH_{\omega}}f= f,\;t \in \IR \}$ is one dimensional, $E_{\omega}\pi_{\omega}(\IM)''E_{\omega}$ is one dimensional and thus in particular abelian. By Proposition 4.3.7 and Theorem 5.3.37 in [BR], $\omega$ is a factor state if and only if $\omega$ is a pure state. Thus by Corollary 1.3, $\omega$ is not a factor state of $\IM$. 

\vsp 
Suppose $H_{\omega}$ has a mass gap. Then by Theorem 2 in [NaS], we verify R. T. Powers criteria [Po] (19) for factor property of $\omega$. This completes the proof for (b) by (a). 
\end{proof} 
 
\vsp
\NI {\bf XY model: } We consider the exactly solvable XY model. The Hamiltonian $H_{XY}$ of the XY model is
determined by the following prescription:

$$H_{XY}= J ( \sum_{j \in \IZ} \{ \sigma_x^{(j)}\sigma_x^{(j+1)}+\sigma_y^{(j)} \sigma_y^{(j+1)} \} - 
2 \lambda \sum_{j \in \IZ} \sigma_z^{(j)}),$$
where $\lambda$ is a real parameter stand for external magnetic field and $J$ is a non-zero real number, 
$\sigma_x^{(j)},\sigma_y^{(j)}$ and $\sigma_z^{(j)}$ are Pauli spin matrices at site $j$. It is well known [AMa] that ground state exists and is unique. It is simple to verify that $\tilde{H}=H$ since we can rewrite 
$H_{XY}$ as sum over elements of the form $\sigma_x^{(j-1)} \sigma_x^{(j)} + \sigma_y^{(j-1)} \sigma_y^{(j)}$. 
Since the transpose of $\sigma_x$ is itself, transpose of $\sigma_y$ is $-\sigma_y$ and transpose of 
$\sigma_z$ is itself, we also verify that $H_{XY}^t=H_{XY}$. Hence $H_{XY}$ is real and refection symmetric.  

\vsp 
For $J < 0$, it is also well known that for $|\lambda| \ge 1$, the unique ground state is a product state thus split state. On the other hand for $|\lambda| < 1$, the unique ground state is not a split state [Ma2 Theorem 4.3]. For $J > 0$, $H_{XY}$ is reflection symmetric with a twist $r_0$ which rotates an angle $\pi$ with respect to $Y$-axis. Furthermore, by a general theorem [FILS] $\omega$ is also reflection positive with a twist $r_0=\sigma_y$ for $J > 0$ and $\lambda =0$. Thus by Corollary 4.5, the unique ground state of $H_{XY}$ model is a non split state and $H$ is not a subset of $\{1,-1 \}$. In such a case, a simple application of Theorem 1.3 in [Mo4] says that the spacial correlation functions of the ground state does not decay exponentially. 
\qed       

\vsp
\NI {\bf XXX MODEL: } Here we consider the prime example where very few exact results on its ground state are known. The Hamiltonian $H_{XXX}$ of the spin $s$ anti-ferromagnetic chain i.e. the Heisenberg XXX model is determined by the following formula:
$$H_{XXX} = J \sum_{j \in \IZ} \{ S_x^{(j)}S_x^{(j+1)}+S_y^{(j)}S_y^{(j+1)} + 
S_z^{(j)} S_z^{(j+1)} \}$$
where $S_x^{(j)},S_y^{(j)}$ and $S_z^{(j)}$ are representation in $d=2s+1$ dimensional of Pauli spin matrices 
$\sigma_x,\sigma_y$ and $\sigma_z$ respectively at site $j$. 
Since $H_{XXX}$ can be rewritten as sum of elements of the form  
$$\{ S_x^{(j-1)}S_x^{(j)}+S_y^{(j-1)}S_y^{(j)} +  S_z^{(j-1)} S_z^{(j)}\},$$
it is simple to check that $\gamma(\tilde{H}_{XXX})=H_{XXX}$. 

We also claim that 
$$H_{XXX}^t=H_{XXX}$$ 
To that end, we consider the space $V_d$ of homogeneous polynomials in two
complex variable with degree $d,\;d \ge 2$ i.e. $V_d$ is the space of functions of the form
$$f(z_1,z_2)=a_0z^d_1+a_1z_1^{d-1}z_2+...+a_dz_2^d $$ 
with $z_1,z_2 \in \IC$ and $a_i's $ are arbitrary complex constants. Thus $V_d$ is a $d$-dimensional complex vector space. The $d-$dimensional irreducible representation 
$\pi_d$ of the Lie-algebra $su_2(\IC)$ is given by 
$$\pi_d(X)f= - { \partial f \over \partial z_1 } ( X_{11}z_1+X_{12}z_2 ) +  
{ \partial f \over \partial z_2 } ( X_{21}z_1+X_{22}z_2 ),$$  
where $X$ in any element in Lie-algebra $su_2(\IC)$. It is simple to verify that the transpose of $S_x=\pi_d(\sigma_x)$ is itself, transpose of $S_y=\pi_d(\sigma_y)$ is $-S_y$ and transpose of $S_z=\pi_d(\sigma_z)$ is itself. Thus $H_{XXX}^t=H_{XXX}$ for any $d=2s+1$. Furthermore, for $J > 0$, the unique positive temperature state (KMS state ) for $H_{XXX}$ is also reflection positive [FILS] with twist $r_0$ since $H_{XXX}$ admits the functional form (14) [FILS]. Thus any limit point of KMS states as temperature goes to zero is also reflection positive for $J > 0$. 

\vsp 
Thus for $J \neq 0$, if ground state of $H_{XXX}$ is unique then the unique ground state is also pure, real and reflection positive with twist $r_0$. This brings a contradiction for even values of $d$ by Theorem 4.5. Thus in particular, for $J > 0$ ( anti-ferromagnet ) and half-odd integer $s$, any low temperature limit point ground state of temperature states is not pure. Same holds good for ferromagnet model as well which is not surprising.  

\vsp 
An interesting point here to be noted that these pure ground states are not limit points of their temperature states. So we conclude that there are solutions for ground states for $H_{XXX}$ other than the well known infinite volume Bethe Ansatz [Be] solution for even values of $d$. 

\vsp 
As an indirect consequence, we conclude that the Bethe solution is not pure as follows: The symmetrized finite truncation of $H_{XXX}$ i.e.
$$H_{XXX}^n = \sum_{-n+1 \le k \le n } \theta^k(h^{XXX}_0)$$ with periodic 
boundary condition admits the functional form of (14) and thus each such truncated Hamiltonian has a unique temperature state at inverse temperature $\beta$ and the state is refection positive with twist $r_0$. Thus the unique ground state of $H_{XXX}^n$ [AL] is also reflection positive with twist $r_0$. This shows an infinite volume limit of ground states of symmetrized truncated Hamiltonians is also reflection positive with twist $r_0$. That a limiting infinite volume state is also translation and $SU(2)$ invariant follows along the same line of argument. It proves that an infinite volume limit point of Bethe states is not pure by Theorem 4.5 for even values of $d$. The same statement holds true for even values of $d$ if we deal with quasi lattice $H_{XXX}$ models [Ma3] with periodic boundary condition and odd many legs.    

\bigskip
{\centerline {\bf REFERENCES}}

\begin{itemize} 
\bigskip
\item{[Ac]} Accardi, L. : A non-commutative Markov property, (in Russian), Functional.  anal. i Prilozen 9, 1-8 (1975).

\item{[AC]} Accardi, Luigi; Cecchini, Carlo: Conditional expectations in von Neumann algebras and a theorem of Takesaki.
J. Funct. Anal. 45 (1982), no. 2, 245–273.

\item{[Ae]} Aernout, van enter: A private communication, June, 2013.  

\item{[AKLT]} Affleck, L.; Kennedy, T.; Lieb, E.H.; Tasaki, H.: Valence Bond States in Isotropic Quantum Antiferromagnets, Commun. Math. Phys. 115, 477-528 (1988). 

\item{[AL]} Affleck, L.; Lieb, E.H.: A Proof of Part of Haldane's Conjecture on Spin Chains, Lett. Math. Phys, 12,
57-69 (1986). 

\item{[Ara1]} Araki, H.:  Gibbs states of a one dimensional quantum lattice. Comm. Math. Phys. 14 120-157 (1969). 

\item{[Ara2]} Araki, H.: On uniqueness of KMS-states of one-dimensional quantum lattice systems, Comm. Maths. Phys. 44, 1-7 (1975).

\item{[AMa]} Araki, H.; Matsui, T.: Ground states of the XY model, Commun. Math. Phys. 101, 213-245 (1985).

\item{[Ar]} W.B. Arveson, Continuous analogues of Fock space I, Mem. Amer. Math. Soc. 80
(1989), no. 409.

\item{[Be]} Bethe, H.: Zur Theorie der Metalle. I. Eigenwerte und Eigenfunktionen der linearen Atomkette. (On the theory of metals. I. Eigenvalues and eigenfunctions of the linear atom chain), 
Zeitschrift für Physik A, Vol. 71, pp. 205-226 (1931).

\item{[BR-I]} Bratteli, Ola,; Robinson, D.W. : Operator algebras
and quantum statistical mechanics, I, Springer 1981.

\item{[BR-II]} Bratteli, Ola,; Robinson, D.W. : Operator algebras
and quantum statistical mechanics, II, Springer 1981.

\item{[BE]} Bratteli, Ola; Evans, David E. Evans: Dynamical semigroups commuting with
compact abelian actions. Ergodic Theory and Dynamical Systems, 3, pp 187-217 (1983)

\item{[BJP]} Bratteli, Ola,; Jorgensen, Palle E.T. and Price, G.L.: 
Endomorphism of $\clb(\clh)$, Quantisation, nonlinear partial differential 
equations, Operator algebras, ( Cambridge, MA, 1994), 93-138, Proc. Sympos.
Pure Math 59, Amer. Math. Soc. Providence, RT 1996.

\item{[BJKW]} Bratteli, Ola,; Jorgensen, Palle E.T.; Kishimoto, Akitaka and
Werner Reinhard F.: Pure states on $\clo_d$, J.Operator Theory 43 (2000),
no-1, 97-143.    

\item{[BJ]} Bratteli, Ola ; Jorgensen, Palle E.T. :Endomorphism of $\clb(\clh)$, II, 
Finitely correlated states on $\clo_N$, J. Functional Analysis 145, 323-373 (1997). 

\item{[Cun]} Cuntz, J.: Simple $C^*$-algebras generated by isometries. Comm. Math. Phys. 57, 
no. 2, 173--185 (1977).  

\item{[DR]} Dagotto, E.;Rice, T.M.: Surprise on the way from one to two dimensional quantum magnets: The ladder materials, Sciences 271, 618-623 (1996)  

\item{[DLS]} Dyson, Freeman J.; Lieb, Elliott H.; Simon, B.: Barry Phase transitions in quantum spin systems with isotropic and non-isotropic interactions. J. Statistical. Phys. 18 (1978), no. 4, 335-383.

\item{[Ef]} Efstratios Manousakis: The spin-$1 \over 2$ Heisenberg antiferromagnet on a square lattice and its application to the cuprous oxides, Rev. Mod. Phys. 63, 1–62 (1991)

\item{[Ev]} Evans, David E.: Irreducible quantum dynamical
semigroups, Commun. Math. Phys. 54, 293-297 (1977).

\item{[EvK]} Evans, David E.; Kawahigashi, Yasuyuki: Quantum symmetries on operator algebras. Oxford Mathematical Monographs. 
Oxford Science Publications. The Clarendon Press, Oxford University Press, New York, 1998.  

\item{[Fa]} Faddeev, L.D.: Algebraic Aspects of Bethe-Ansatz, Int.J.Mod.Phys. A10 (1995) 1845-1878. 

\item{[FNW1]} Fannes, M.; Nachtergaele, B.; Werner, R.: Finitely correlated states on quantum spin chains
Commun. Math. Phys. 144, 443-490(1992).

\item{[FNW2]} Fannes, M.; Nachtergaele, B.; Werner, R.: Finitely correlated pure states, J. Funct. Anal. 120, 511-
534 (1994). 

\item{[FNW3]} Fannes, M.; Nachtergaele B.; Werner, R.: Abundance of translation invariant states on quantum spin chains, Lett. Math. Phys. 25 no.3, 249-258 (1992).

\item{[Fr]} Frigerio, A.: Stationary states of quantum dynamical semigroups, Comm. Math. Phys. 63 (1978) 269-276.

\item{[FILS]} Fr\"{o}hlich, J.; Israel, R., Lieb, E.H., Simon, B.: Phase Transitions and Reflection Positivity-I 
general theory and long range lattice models, Comm. Math. Phys. 62 (1978), 1-34 

\item{[GM]} Ghosh, D.; Majumdar, C.K.: On Next ‐ Nearest ‐ Neighbour Interaction in Linear Chain. J. Math. Phys, 10, 1388 (1969). 

\item{[Hag]} Haag, R.: Local quantum physics, Fields, Particles, Algebras, Springer 1992. 

\item{[Ha]} Hall, B.C. : Lie groups, Lie algebras and representations: an elementary introduction, Springer 2003. 

\item {[Ka]} Kadison, Richard V.: Isometries of operator algebras, Ann. Math. 54(2)(1951) 325-338.  

\item{[Ki]} Kishimoto, A.: On uniqueness of KMS-states of one-dimensional quantum lattice systems, Comm. Maths. Phys. 47, 167-170 (1976).

\item{[Kir]} Kirillov, A. A.: Elements of the theory of representations. Springer-Verlag, Berlin-New York, 1976.

\item{[LSM]} Lieb, L.; Schultz,T.; Mattis, D.: Two soluble models of an anti-ferromagnetic chain Ann. Phys. (N.Y.) 16, 407-466 (1961).

\item{[Ma1]} Matsui, A.: Ground states of fermions on lattices, Comm. Math. Phys. 182, no.3 723-751 (1996). 

\item{[Ma2]} Matsui, T.: A characterization of pure finitely correlated states. 
Infin. Dimens. Anal. Quantum Probab. Relat. Top. 1, no. 4, 647--661 (1998).

\item{[Ma3]} Matsui, T.: The split property and the symmetry breaking of the quantum spin chain, Comm. 
Maths. Phys vol-218, 293-416 (2001) 

\item{[Ma4]} Matsui, T.: On the absence of non-periodic ground states for the antiferromagnetic XXZ model. Comm. Math. Phys. 253 (2005), no. 3, 585-609.

\item{[Mo1]} Mohari, A.: Pure inductive limit state and Kolmogorov's property. II. J. Operator Theory 72 (2014), no. 2, 387-404.

\item{[Mo2]} Mohari, A.: Translation invariant pure state on $\otimes_{k \in \IZ}\!M^{(k)}_d(\IC)$ and Haag duality, Complex Anal. Oper. Theory 8 (2014), no. 3, 745-789. 

\item{[Mo3]} Mohari, A.: Translation invariant pure state on $\clb=\otimes_{k \in \IZ}M^{(k)}_d(\IC)$ and its split property, J. Math. Phys. 56, 061701 (2015).

\item{[Mo4]} Mohari, A.: Isomorphism theorem for Kolmogorov states of $\IM = \dsp{\otimes_{j \in \IZ}}\!M^{(j)}_d(\IC)$, arXiv:1309.7606  

\item{[Mo5]} Mohari, A.: Translation invariant states and its mean entropy, ??   

\item{[Na]} Nachtergaele, B. Quantum Spin Systems after DLS1978, `` Spectral Theory and Mathematical Physics: A Festschrift in Honor of Barry Simon's 60th Birthday '' Fritz Gesztesy et al. (Eds), 
Proceedings of Symposia in Pure Mathematics, Vol 76, part 1, pp 47--68, AMS, 2007. 

\item{[NaS]} Nachtergaele, B; Sims, R. : Lieb-Robinson bounds and the exponential clustering theorem, Comm. Math. Phys. 265, 119-130 (2006). 

\item{[Po]} Popescu, G.: Isometric dilations for infinite sequences of non-commutating operators, Trans. Amer. Math.
Soc. 316 no-2, 523-536 (1989)

\item{[Pow]} Powers, Robert T.: Representations of uniformly hyper-finite algebras and their associated von Neumann. rings, Annals of Math. 86 (1967), 138-171.

\item{[Ru]} Ruelle, D. : Statistical Mechanics, Benjamin, New York-Amsterdam (1969) . 

\item{[Sim]} Simon, B.: The statistical mechanics of lattice gases, vol-1, Princeton series in physics (1993). 

\item{[Sa]} Sakai, S. : Operator algebras in dynamical systems. The theory of unbounded derivations in $C\sp *$-algebras. 
Encyclopedia of Mathematics and its Applications, 41. Cambridge University Press, Cambridge, 1991. 

\item{[S\o]} St\o rmer E.: On projection maps of von Neumann algebras, Math. Scand. 30, 46-50 (1972). 

\item{[Ta1]} Takesaki, M.: Conditional Expectations in von Neumann Algebras, J. Funct. Anal., 9, pp. 306-321 (1972)

\item{[Ta2]} Takesaki, M. : Theory of Operator algebras II, Springer, 2001.

\item{[Wa]} Wassermann, Antony : Ergodic actions of compact groups on operator algebras. I. General theory. Ann. of Math. (2) 130 (1989), no. 2, 273-319

\end{itemize}

\end{document}